\documentclass{aastex}
\usepackage{graphicx,epsfig,emulateapj5}

\newcommand{\etal}{{et al.~}}

\newcommand{\bq}{\begin{equation}}
\newcommand{\eq}{\end{equation}}

\def\gtsim{\lower.5ex\hbox{$\buSildrel > \over\sim$}}
\def\ltsim{\lower.5ex\hbox{$\buildrel < \over\sim$}}
\def\arcsec{^{\prime\prime}}
\def\arcmin{^\prime}

\def\farcs{\hbox{$.\!\!^{\prime\prime}$}}

\def\sun{\mbox{$_\odot$}}
\def\apjl{ApJL}
\def\apj{ApJ}
\def\apjs{ApJS}
\def\mnras{MNRAS}
\def\araa{ARAA}
\def\aj{AJ}
\def\aap{A\&A}
\def\aaps{A\&A Suppl.}

\begin{document}
\title
{Characterizing Bars at  $z\sim$~0 in the optical and NIR: Implications 
for the Evolution of Barred Disks with Redshift}
\author {Irina Marinova\altaffilmark{1}, Shardha Jogee\altaffilmark{1}}
\authoremail{marinova@astro.as.utexas.edu, sj@astro.as.utexas.edu}
\altaffiltext{1}{
Department of Astronomy, University of Texas at Austin, 1 
University Station C1400, Austin, TX 78712-0259}

\begin{abstract}
Critical insights on galaxy evolution stem from the study of bars.
With the advent of high redshift $HST$  surveys 
that trace bars in the rest-frame optical band out to 
$z\sim$~1, it becomes increasingly important to provide a reference baseline 
for bars at $z\sim$~0 in the optical band. 
We present results on bars at $z\sim$~0 in the optical 
and  near-infrared bands, based on 180 spirals in the 
OSUBSGS survey.
(1)  The deprojected  bar fraction at $z\sim$~0 is 
$f_{\rm NIR1}$~$\sim$~60\%~$\pm$~6\% in the near-infrared $H$ band, and 
$f_{\rm optical1}$$\sim$~44\%~$\pm$~6\%  in the optical $B$-band images.
The latter likely miss  bars obscured by dust and star formation. 
(2)~The results before and after deprojection are similar, 
which is encouraging for high redshift studies that forego 
deprojection. 
(3) Studies of bars at $z \sim$~0.2--1.0  (lookback times of 3--8 Gyr)
have reported an optical bar fraction of 
 $f_{\rm optical2}$$\sim$~30\% $\pm$ 6\%, after applying cutoffs in absolute 
magnitude  ($M_{\rm V}<$-19.3), bar size ($a_{\rm bar} \ge$ 1.5 kpc), and 
bar ellipticity ($e_{\rm bar} \ge$~0.4)  in order to ensure a complete 
sample, adequate spatial resolution, and reliable bar identification 
out to $z\sim$~1.  
Applying these exact cutoffs in magnitude,  bar size, and bar ellipticity 
to the  OSUBSGS data yields  a comparable  optical $B$-band bar fraction 
at $z\sim$~0 of $f_{\rm optical3}$$\sim$~34\%~$\pm$~6\%.
This rules out scenarios where the optical bar fraction in bright 
disks declines strongly with redshift.
(4) We investigate bar strengths at  $z \sim$~0 using the maximum 
bar ellipticity ($e_{\rm bar}$) as a guide.
Most  ($\sim$~70\%) bars have moderate to high ellipticity 
(0.50~$ \le  e_{\rm bar} \le$~0.75), and only a  small  
fraction (7\%--10\%)  have 0.25~$ \le  e_{\rm bar} \le$~0.40.
There is no bimodality in the distribution of $e_{\rm bar}$.
The $H$-band bar fraction and  $e_{\rm bar}$ show no substantial 
variation across RC3 Hubble types Sa to Scd.
(5)~RC3 bar types should be used with caution. 
Many  galaxies with RC3 types `AB' 
turn out to be unbarred and RC3 bar classes  `B' and `AB'  
have a significant overlap in  $e_{\rm bar}$.
(6)~Most  (68\% in $B$ and 76\% in $H$)  bars have sizes  below 5 kpc.
Bar and disk sizes correlate, and the ratio ($a_{\rm bar}$/$R_{\rm 25}$) 
lies primarily in the range 0.1 to 0.5. 
This suggests that the growth of bars and disks is intimately tied.
\end {abstract}


\section{Introduction}

Stellar bars  are recognized as the  most important internal 
factor that  redistributes the angular momentum of
the baryonic and dark matter  components of  disk galaxies
(e.g., Weinberg 1985; Debattista \& Sellwood 1998, 2000; 
Athanassoula 2002; Berentzen, Shlosman, \& Jogee 2006),  
thereby driving their dynamical and secular evolution. 
Bars efficiently drive gas from the outer disk to the 
central few hundred parsecs  and  are observed to feed 
central starbursts in local galaxies (Elmegreen 1994; Knapen 
\etal 1995; Hunt \& Malakan 1999; Jogee \etal 1999;  Jogee, 
Scoville, \& Kenney 2005). 
It remains a matter of contention whether large-scale bars relate to 
AGN activity in galaxies, given the reduction by several orders of 
magnitude needed in the specific angular momentum of gas before it 
can feed a central black hole, and conflicting observational results 
(see review by Jogee 2006 and references therein;  also Mulchaey 
\& Regan 1997; Knapen \etal  2000; Laine \etal 2002; Laurikainen \etal 2004). 
In several galaxies, bar-driven gas inflows appear intimately 
tied to the formation of disky, high $v/\sigma$  stellar components 
in the inner kpc, or `pseudobulges' (Kormendy 1993; Jogee 1999; review by Kormendy \& Kennicutt 2004; Jogee, Scoville, \& Kenney 2005; Athanassoula 2005). Furthermore,  
the orbital structure of bars can lead to the observed peanut-shaped 
and boxy bulges in inclined galaxies (Combes \etal 1990; Pfenniger 
\& Norman 1990;  Bureau \& Athanassoula 2005; 
Athanassoula 2005; Martinez-Valpuesta \etal 2006; 
Debattista \etal 2006).

Earlier ${\it Hubble~Space~Telescope~(HST)}$  studies at optical 
wavelengths (e.g., Abraham \etal 1999) reported a paucity of 
stellar bars and a sharply declining optical bar fraction 
at intermediate redshifts $z >$~0.5. 
Studies at near-infrared (NIR) wavelengths also found a low 
bar fraction, but the authors rightly concluded that 
the large effective point spread functions (PSFs) of the 
NIR camera only allowed the detection of  large bars whose  
semi-major axes  exceeded  $0.9\arcsec$, corresponding to 
7.2 kpc\footnote{We assume in this
paper a flat cosmology with $\Omega_M = 1 - \Omega_{\Lambda} = 0.3$
and $H_{\rm 0}$ =70~km~s$^{-1}$~Mpc$^{-1}$.}
at $z\sim$~1.0 (Sheth \etal 2003).
Recent works based on large optical surveys have  now demonstrated 
the abundance of bars  at intermediate redshifts $z \sim$~0.2--1.0, 
corresponding to lookback times of  3--8 Gyr 
(Elmegreen \etal 2004; Jogee \etal 2004;  Zheng \etal  2005; Sheth \etal in preparation).
The fundamental issue of how robust bars are, and the associated 
implications for bar-driven  evolution in disks  over the last 
10 Gyr, remains open (e.g., Jogee \etal  2004; Shen \& Sellwood 2004; Athanassoula, Lambert, \& Dehnen  2005;  Bournaud \etal 2005; Berentzen, Shlosman, 
\& Jogee 2006; Berentzen \& Shlosman 2006; 
Martinez-Valpuesta \etal 2006; Debattista \etal 2006).

In order to put bars in a cosmological context, it now behooves us 
to characterize the frequency  and impact of bars 
by applying the {\it same quantitative} methods to large samples 
at $z\sim$~0 and at higher redshifts. 
Spurred by these considerations, we characterize in this paper 
the frequency and structural properties of bars in the 
local Universe at optical and NIR  wavelengths, by 
ellipse-fitting the $B$ and $H$ images of the OSU Bright Spiral 
Galaxy Survey (OSUBSGS; Eskridge et al. 2002) of 180 spirals. 
The first goal of  this study is   to  provide 
{\it quantitative} 
characterizations  of the bar fraction $f_{\rm bar}$  (defined 
as the fraction of disk galaxies that are barred) and structural 
properties (sizes, ellipticities, etc.) of bars  at 
$z\sim$~0,  as a function of wavelength, Hubble types, 
and host galaxy properties. 
Furthermore, with the advent of high redshift $HST$ surveys, such as 
the  Tadpole field (Tran \etal 2003), 
the Galaxy Evolution  from Morphology and SEDs (GEMS; Rix \etal 2004), 
the Great Observatories Origins Deep Survey (GOODS; Giavalisco \etal 
2004), and COSMOS (Scoville \etal 2006), which  
trace bars in the rest-frame optical band out to $z\sim$~1,
it becomes increasingly important to provide a reference baseline
for bars at $z\sim$~0 in the optical band.
Thus, a second  goal of our study is to provide {\it a rest-frame 
optical }  $z \sim$~0 point for bars  based on  ellipse fits, in order
to directly compare  with studies of intermediate-redshift bars  
(Jogee \etal  2004; Elmegreen \etal 2004; Zheng \etal  2005; Sheth et al. 
in preparation) that also use ellipse fits.
In particular, we use in this paper the same procedure of ellipse 
fits ($\S$~3.1) and the same quantitative  characterizations ($\S$~3.3)
of bars that were  applied by Jogee \etal  (2004) to bars 
at intermediate redshifts ($z \sim$~0.2--1.0) 
in  the GEMS survey.

Several studies have used the OSUBSGS to  gauge bars in the local 
Universe (e.g., Eskridge et al. 2000; Block et al. 2002; Whyte \etal 2002; Buta \etal 
2005), but they  differ significantly from our study and cannot meet 
our two goals. Eskridge \etal (2000) visually classified bars in 
the $H$ band, and in the $B$ band, they used the Third Reference 
Catalog of Bright Galaxies (de Vaucouleurs et al. 1991; hereafter RC3) 
visual  bar classes. Such visual classifications form an invaluable 
first step, but by definition, are subjective and difficult to 
compare with results from other studies. Block \etal (2002) and later Buta \etal (2005)  applied 
the gravitational torque $Q_{\rm b}$ method, based on Fourier amplitudes, 
to $H$-band images of 163 and 147 OSUBSGS galaxies, respectively. This quantitative method 
is less subjective than visual classification, but the results of
Block \etal (2002) and Buta \etal (2005) cannot be compared to intermediate redshift studies 
for two reasons. First, the latter studies were based on the 
{\it HST} Advanced Camera for Surveys (ACS) data and trace the 
rest-frame optical properties of bars, while Block \etal (2002) and 
Buta \etal (2005) 
deal with the rest-frame NIR. Second, it is non-trivial to 
derive  $Q_{\rm b}$ for intermediate-redshift galaxies  
because of resolution and signal-to-noise limitations.  Whyte 
\etal (2002) fitted ellipses to $B$-band images of only 89 of the 
180 OSUBSGS galaxies,
and do not provide a distribution of bar properties
as a function of Hubble type. Our 
present study 
complements these existing studies by ellipse fitting $B$-band and 
$H$-band images of  all  180 OSUBSGS galaxies, and performing a 
comprehensive, statistically significant analysis of  barred galaxies 
in the local Universe. 
It complements the ongoing analysis (Barazza, Jogee, \& Marinova 2006) \
of local bars based on a sample of 5000 galaxies in the Sloan Digitized Sky 
Survey (SDSS).

The outline of this paper is as follows. $\S$~2 discusses  
the sample selection based on the OSUBSGS survey (Eskridge \etal 2002).
$\S$~3 describes  the ellipse-fitting method, the criteria 
used for identifying bars, and  deprojection of images 
and profiles to face-on.
In  $\S$~4.1--4.4, we present results on 
the  bar fraction at $z\sim$~0,   
its dependence on Hubble type, the distribution of bar sizes and 
strengths as characterized by ellipse-fitting, and  
the variation of bar properties along the  
Hubble sequence. Results  are presented both before and after 
deprojection to face-on. 
In  $\S$~4.5, we present a first-order comparison of the bar 
fraction and properties at $z\sim$~0 from OSUBSGS to those derived
at  $z\sim$~0.2--1.0 or lookback times of  3--8 Gyr  
from GEMS (Jogee \etal  2004) and the Tadpole field (Elmegreen \etal 2004). 
In $\S$~4.6, we discuss the constraints set by our results 
for theoretical models addressing the  robustness of bars, 
and  the assembly of the Hubble sequence over cosmological times.
$\S$ 5 presents the summary and conclusions.

This paper is the first in a series of three based on the OSUBSGS.
Paper II (Marinova \etal in prep) will address the bulge 
properties and activity of barred and unbarred galaxies in the 
OSUBSGS sample. In paper III, we will present simulations 
that artificially redshift the  rest-frame optical and NIR images 
of the local OSUBSGS sample out to   $z\sim$~1--2, in order to 
assess the impact of redshift-dependent systematic effects 
on the recovery rate of bars  in surveys conducted 
by current and future facilities in the optical and IR,
such as the planned Wide Field Camera 3 (WFC3) and the James
Webb Space Telescopes (JWST).

\section{Data and Sample}

The OSUBSGS targets local spiral
galaxies that  are taken from the RC3 catalog 
and  chosen to represent the bright disk galaxy
population in the local universe (Eskridge \etal 2002). 
The galaxies are selected using the following 
criteria: RC3 type of S0/a or later, (0$\leq$ T $\leq$9), $M_{B}< 12$, $
D_{25}< 6\arcmin.5$, and $-80^\circ <\delta< +50^\circ$
(Eskridge et al. 2002), and are  imaged in the $B$, $V$, $R$, $H$, $J$, 
and $K$ bands. The  $B$ and $H$ images of 182 OSUBSGS galaxies 
are available as part of a public data release  (Eskridge \etal 2002).
Our starting sample (sample S1) consists of the afore-mentioned 
182 OSUBSGS galaxies with $B$ and/or $H$ images.
After discarding galaxies (2 galaxies or 1\% of sample S1) that do not have 
images in both the $B$ and $H$ bands, we are left with sample S2 of 180 
galaxies imaged in both bands. This constitutes the sample of galaxies 
to which we fitted ellipses in order to characterize bars and disks, 
as outlined in $\S$ 3.

\section{Method for Characterizing Bars and Disks}

We adopt the widely used procedure  of characterizing bars
 and disks in galaxies via ellipse fits  
(e.g., Wozniak \etal 
 1995, Friedli \etal 1996; Regan \etal 1997; Mulchaey \& Regan 1997; 
 Jogee \etal  1999, 2002a,b, 2004; Knapen \etal 2000; Laine \etal 2002;    
 Sheth \etal 2003; Elmegreen \etal 2004), as described in detail 
in $\S$~3.1. 
Our analysis procedure is schematically illustrated in Figure 1
and described in  sections 3.1 to 3.4.

\subsection{Ellipse Fitting}

We start with the sample S2 of 180 galaxies imaged in both the $B$ and 
$H$ bands (Fig. 1). 
We first remove stars from  the $B$- and  $H$-band images 
of each galaxy by replacing them with the average of the sky
background using a circular aperture. We then find the center of the 
galaxy using  the IRAF
routine `imcenter'. We determine a maximum galaxy semi-major axis 
length ($a_{\rm max}$)  out to which ellipses will be fitted in each 
image by finding out  where the galaxy isophotes reach the sky level. 

We  then use the standard IRAF task `ellipse' to fit ellipses  to 
each image out to $a_{\rm max}$. 
We employ an iterative  wrapper developed by Jogee \etal  (2004)  to 
run `ellipse' up to to 300  times for each object  in order to get 
a good  fit across the whole galaxy. A successful fit is one where 
the routine is able to fit
a ellipse at each radial increment from the center until
it reaches $a_{\rm max}$.  
When using the iraf task 'ellipse' for ellipse fits, the goodness 
of the best fit is measured by four harmonic amplitudes (A3, A4,
B3, B4), which describe by how much the actual isophote differs 
from the best-fitting ellipse  (e.g., Jedrzejewski 1987). We 
have inspected plots of these residuals for representative 
strongly and weakly  barred galaxies (e.g., NGC 4314, 
NGC 613, NGC 1187, NGC 0210, NGC~1300, NGC~7479, 
NGC~5701, NGC~4643, NGC~4548, NGC~4450, NGC~3681, 
NGC~3275, NGC~1703, and NGC~1358). We find that 
the  A3 and B3 residuals are small, typically on the order of 
a few percent. Values for the A4 and B4 residuals typically 
range from  2\% to 10\%, and  do not exceed ~15\%. 

From the final fit for each galaxy, we generate 
radial profiles of surface brightness (SB), ellipticity ($e$), and
position angle (PA).
The fitted ellipses are over-plotted onto the galaxy images to generate 
overlays. Examples of the radial plots and overlays are shown in Figures 
2, 4, and 5.
For each galaxy, an interactive  visualization tool  (Jogee \etal  2004) 
is used to display both the radial profile and the overlays in order 
to perform an extra inspection of the fits. 

Of the 180 galaxies in sample S2, 179 (99\%) and 169 (94\%)  
were successfully fitted in the $H$ and $B$ band, respectively.  
Of the 11 galaxies that could not be fitted in the $B$ band, five 
had strong 
morphological distortions and seem to be interacting; one had a very bright,
saturated star with leakage; and  five had no clearly defined center. The
latter
five galaxies were all of later Hubble type (Sbc and Sc), and had very 
flat or irregular surface brightness profiles in the $B$ band. 
Further analyses to  
characterize inclined, unbarred, and barred disks in $\S$ 3.2 were 
then restricted to the sample S3 of 169 galaxies with successful fits 
in both the $B$ and $H$ bands (Fig. 1).

\subsection{Identifying and excluding highly inclined spirals}

For sample S3, we use the $B$-band images, rather than the $H$-band images, to 
identify and characterize the outer disk  because  
the former are deeper and  trace the disk farther out.
From  the radial profiles generated by 
ellipse-fitting  the $B$-band image, we  measure  the 
ellipticity ($e_{\rm disk}$) and PA  (PA$_{\rm disk}$) 
of the outer disk. The outer disk inclination, $i$, is 
derived from  $e_{\rm disk}$ using cos($i$)~=~(1~-~$e_{\rm disk}$). 
Of the 169 galaxies in sample S3, we find 33 (20\%) 
galaxies  with  disk  inclination $i>60^\circ$ and classify 
them as `inclined'.  They  are listed in the lower part of Table 1. 
Figure 2 shows an example of the $B$-band radial profile and ellipse 
overlays for an inclined galaxy. 

We only use the final sample S4 (Fig. 1) of 136
moderately inclined  ($i <$~60$^\circ$) spirals to further 
characterize the properties of bars (e.g., size, ellipticity, 
frequency)  and disks in $\S$ 3.3--3.4.  Such an
inclination cutoff is routinely applied in morphological
studies because projection effects make it very difficult to
reliably trace structural features in a galaxy that is close
to edge-on.  
The exclusion of highly inclined galaxies does not bias the distribution 
of Hubble types, as shown in Figure 3a, where the Hubble types of samples S3 
and S4 are compared. 
The  absolute $V$-band  magnitudes ($M_{\rm V}$)  of both sample S3
and S4 cover the range -18 to -23, with  most galaxies lying in 
the range $M_{\rm V} \sim$~-20 to -22 (Fig. 3b).

\subsection{Characterizing bars and disks before deprojection}

In $\S$ 3.4, we use the {\it deprojected}  
radial profiles of (SB, $e$, PA) 
to characterize the intrinsic properties of  bars and disks
in sample S4. However, we also decide to first perform the analysis  
on  the {\it observed}  radial profiles {\it before} 
deprojecting them to face-on. 
There are several reasons for this dual approach of 
deriving bar properties both before and after deprojection.
First,  it is useful to have bar properties (e.g., frequency,
strength as characterized by  ellipse-fitting, size) 
prior to deprojection to compare directly to 
studies at intermediate redshifts (Jogee \etal  2004, 
Elmegreen \etal 2004, Zheng \etal  2005), where 
deprojection is not done for several reasons, including the
difficulty in accurately measuring the PA of the line of nodes and  
the inclination of the outer disk in noisy images of distant galaxies.
Second, by having bar properties both before and 
after deprojection, we are able to  assess whether deprojection 
makes a substantial difference to the statistical distributions of 
bar properties. A large difference would raise concerns for intermediate
redshift studies or even for large nearby studies  where deprojection
is often not carried out.

For sample S4, we use the observed radial profiles of (SB, $e$, PA) 
and   the ellipse overlays  to classify galaxies  as `unbarred' 
(Fig. 4)  or `barred' (Fig. 5),  according to the following 
quantitative  criteria.   A galaxy is classified as barred if
the radial variation  of  ellipticity and PA  follows the 
behavior that is expected based on the dominant orbits of a 
barred potential.  Specifically the  following conditions 
must be satisfied before a galaxy is deemed to be barred: 
(1)~The ellipticity, $e$,
increases steadily to a global maximum, $e_{\rm bar}$, greater
than 0.25, while the PA value remains constant (within 10$^\circ$).
This criterion is based on the fact that the main bar-supporting 
orbits, namely the `$x_{\rm1}$' family  of orbits, can be modeled
by concentric ellipses with a constant PA  as a function of radius
in the bar region (Athanassoula 1992a). 
The requirement that the PA must remain constant in the bar 
region is important for excluding other spurious elliptical 
features that may mimic a bar signature in their ellipticity 
profile. 
(2)~Then, at the transition from the bar to the disk region, 
the ellipticity, $e$,  must drop by at least 0.1, and the PA 
usually changes.  
This criterion is justified by the fact that we expect 
a  transition from the highly  
eccentric $x_{\rm1}$ orbits near the bar end to the more 
circular orbits in the disk. We also note  that the drop in
ellipticity by 0.1 at the transition from bar to disk  has 
been shown to work well in identifying bars (e.g., 
Knapen \etal 2000; Laine \etal  2002;  Jogee \etal 2002a, 2002b, 
2004). 

What are the limitations of criteria (1) and (2) in 
identifying bars?
We note that the `constant PA' criterion that we use 
to identify bars may cause us to  miss some weak bars at 
optical wavelengths due to the following reason.
In weak bars, the shock loci and corresponding  dust lanes  
on the leading edge of the bar  are curved (Athanassoula 1992b).
In optical images of  weak bars, these curved dust lanes may 
cause the PA to twist or vary slightly along the bar, thereby 
preventing the  `constant PA' criterion from being met.
In the case of very strong bars,  the  `constant PA' criterion is
a good one and isophotal twist is not an issue, because such 
bars have strong shocks and straight dust lanes along 
their  leading edges (Athanassoula 1992b). 
In order to gauge how many bars we might be missing because of 
the `constant PA' criterion, we identify galaxies that 
show a PA twist accompanied by an ellipticity maximum. It
turns out that  only a small fraction ($\sim$~7\%) 
of galaxies show this effect.

We also note that criterion (1) requires  the peak ellipticity 
($e_{bar}$) over the PA plateau to be greater than 0.25  before 
we call a feature a bar. We picked 0.25 for the practical reason that 
structures with lower ellipticities are quite round and not always
readily distinguishable from disks. Nonetheless, one may be tempted to 
ask whether we would find more bars if this arbitrary limit of 0.25 
were to be lowered, and whether there is a population of low-ellipticity
(e.g., $e_{\rm bar} \sim$~0.10--0.25) bars that we might miss.
We investigated this question using the OSUBSGS sample, and find that 
there is no increase in the number of bars if the limiting value for 
$e_{\rm bar}$ in criterion (1) were to be lowered from  0.25  to 0.10. 
The reason for this becomes clear later, in Figure 13, which 
shows that the number of bars already starts to drop rapidly for 
ellipticities below 0.40, such that by the time we reach $e_{\rm bar}$ of 
0.25, we are already probing the tail end of bar distributions.

In addition to classifying galaxies as `barred' and `unbarred', 
we also use the radial profiles to derive the structural properties of 
the bar and disk.  Specifically, 
for all galaxies, we measure the  ellipticity, PA, and  
semi-major axis of the outer disk  ($e_{\rm disk}$, PA$_{\rm disk}$,
$a_{\rm disk}$).  For galaxies classified as `barred', we also 
measure the maximum ellipticity  ($e_{bar}$), the PA, and the semi-major 
axis of the bar. We will discuss in  $\S$~4.3 how 
the maximum bar ellipticity  ($e_{\rm bar}$) constrains the  bar 
strength.  Here, we discuss the question of how to locate the end of 
the bar in order to measure  the bar semi-major  axis. 
There has been some discussion in the 
literature as to  whether the bar end should be defined as 
the  radius ($a_{\rm bar}$) where the bar ellipticity is a  
maximum,  or as the radius where the PA changes abruptly at 
the transition from the bar to the disk. 
From a theoretical perspective, several early simulations  
(e.g., Athanassoula 1992a; O'Neill \& Dubinski 2003) show 
that the definition of bar length based on `peak ellipticity' 
can underestimate the true extent of the bar.
Recently, Martinez-Valpuesta, Shlosman, \& Heller (2006) have
performed a systematic study of the radius ($a_{\rm bar}$) 
of maximum bar ellipticity  and the bar length. They show  
that there is a very good correspondence between two 
independent methods to determine the bar size: ellipse fitting and orbital
analysis. The orbital analysis has involved finding the largest (Jacobi) 
energy $x_{\rm1}$ orbit in the bar that is still stable. The ellipse 
fitting becomes better if the size of the bar is given by the radius 
where the ellipticity declines by  15\% from its maximal value.

In his empirical study of bar sizes using ellipse fits, Erwin (2005)
argues that using the PA signature to define the bar size 
provides an upper limit, and that the two measures of bar length
are very well correlated. However, he finds that it is harder to
unambiguously measure the bar size from the PA criterion and that the 
definition of bar size based on peak ellipticity is more readily applied
consistently to a large number of different galaxy morphologies (Erwin
2005).  In this study, we have adopted the first approach. We 
use  the semi  major axis ($a_{\rm bar}$) where the  maximum bar 
ellipticity occurs as  a measure of the bar length.
We caution that this may underestimate the bar length  in some galaxies.
However, a  visual comparison of $a_{\rm bar}$ with the images of 
our galaxies suggests that $a_{\rm bar}$ does a reasonable job in most
cases.

\subsection{Characterizing bars and disks after deprojection}

For sample S4, we use the inclination, $i$, and the PA of the 
outer disk (determined in $\S$ 3.2) to {\it analytically}
deproject  the observed  $H$ and $B$ band  radial profiles of ($e$, PA) 
to face-on. 
We perform the  analytical deprojection using  a code developed by  
Laine et al. (2002) and used previously  in Laine \etal (2002) 
and Jogee \etal (2002a,b). It should be noted that the deprojection formula 
used in the code only strictly applies to infinitesimally thin structures, 
and may be inaccurate near
the galaxy center in the vicinity of the bulge. However, it is a reasonable 
 approximation in the region of interest where large-scale bars reside.
Figure 6 shows an example of the deprojected  
radial profiles of NGC 4548 in the $B$ and $H$ bands overlaid on the 
observed profiles.

We note that the process of analytically deprojecting 
the radial {\it profiles} to face-on after ellipse-fitting the 
observed (i.e, un-deprojected) images is analogous to the 
process of first deprojecting the observed 
{\it images} to face-on, and then ellipse-fitting the deprojected 
images in order to generate face-on radial {\it profiles}. 
The two methods should yield the same results unless the
images are very noisy.
We verified this expectation with the following steps. 
(1) We deproject the {\it images}  of several galaxies using 
the  Multichannel Image Reconstruction, Image Analysis 
and Display (MIRIAD) routine `deproject'. 
The routine takes as input  the observed image, 
the galaxy center, the inclination $i$ and PA of the outer disk,
and outputs the deprojected image; 
(2) We then fit ellipses to these deprojected images using the procedure 
outlined in $\S$~3.1, and generate face-on radial profiles 
of  SB, $e$, and PA;
(3)  These face-on radial profiles generated from the
deprojected images, are compared with the deprojected radial  profiles 
derived analytically  from the  the observed profile.  
There is good agreement in all  cases, showing that we are
not noise limited.

This is illustrated in Figure 7 for the $B$ band image of NGC 4548.
The observed  and deprojected images are shown in the left panel.
In the right panel,  three radial profiles are plotted: 
the observed radial profile derived by fitting ellipses 
to the observed image  is plotted as stars; the deprojected radial 
profile derived analytically from  the observed profile  
is plotted as squares; and the face-on radial profile 
derived by fitting ellipses to the deprojected image 
is plotted as triangles.  There is good agreement 
between the squares and the triangles.

The deprojected profiles provide  an accurate characterization
of the `intrinsic' or face-on properties of disks and bars.
For all galaxies in S4, we therefore use the analytically 
deprojected $B$ and $H$  radial profiles to classify galaxies 
as `barred' or  `unbarred', according 
to the criteria outlined in $\S$ 3.3.   
We also re-measure the bar ellipticity ($e_{bar}$), 
semi-major axis ($a_{bar}$), and  disk size $a_{\rm disk}$ 
 from the deprojected radial profile.
In the rest of this paper, many of these  deprojected quantities 
will be compared to those derived before  deprojection ($\S$ 3.3)
in order to gauge  the impact of deprojection.

\section{Results and Discussions}

\subsection{The optical and NIR bar fraction at $z\sim$~0}

Table 2 and Figure 8 show the bar fraction  (defined as the fraction of 
spiral galaxies that are  barred) for the $B$ and $H$ bands, both before 
($\S$~3.3) and after deprojection  ($\S$~3.4).
The results are based on sample S4 of 136 moderately inclined ($i< 60^\circ$) spirals ($\S$~3.2).
The sample is dominated by galaxies with $M_{\rm V} \sim$~-20 to -22. 
We find a deprojected  bar fraction of 60\%  in the $H$ band 
and a lower fraction of  44\% in the $B$-band images, which likely 
miss  bars obscured by dust and star formation.
Our results that 60\% of spirals are barred in the infrared
confirms  the preponderance of bars among spirals in the local 
Universe. 

Our $H$-band bar fraction of $\sim$~60\% is in agreement with the NIR 
bar fraction of 59\%  (Menendez-Delmestre \etal 2006) based on 2MASS. 
It  is also consistent, within a margin of 12\%, with the results of Eskridge \etal  (2000), 
who visually  inspected  the OSUBSGS  $H$-band  images and reported 
an  overall  $H$-band bar fraction of 72\%,  with 56\% of spirals
hosting `strong' bars and 16\% hosting `weak' bars. 
Why is there a 12\% deviation? The Eskridge \etal  (2000) paper 
does not give `barred' or `unbarred'  classifications for individual
galaxies, so we can not make a case by case comparison with that 
study. However, in a subsequent paper,  Eskridge \etal (2002) give 
visual classifications of  individual galaxies as barred or unbarred, 
and classify barred systems as  `SB'  (strongly barred) and  
`SAB' (weakly barred).  We find that our classifications as barred or
unbarred disagree on 25 galaxies in the $B$ band ($\sim$~18\% of
sample S4), and 23 galaxies in the $H$ band ($\sim$~17\% of sample
S4).  Of the galaxies in the  $B$ band and  $H$ band where we differ, 
we find that the majority (15 of the 25 galaxies in the $B$ band,  and 
11 of the 23  galaxies in the $H$ band) are  classified as `SAB' 
(weakly barred) by Eskridge \etal (2002).  We conclude that, as might be
intuitively expected, the differences between visual and  quantitative
classifications of bars are strongest for systems that
visually  appear as `weakly barred'.

How does our study  compare with other quantitative studies?
We find that our reported  $H$-band bar fraction of 60\%   
agrees with that of Laurikainen \etal (2004),   who used  
Fourier modes and the $Q_{\rm b}$ method for 158 galaxies in the
OSUBSGS sample and 22 2MASS galaxies. Laurikainen \etal  (2004)
find a NIR bar fraction of 62\% for galaxies with $i< 60^\circ$.  
We  present a more detailed comparison of 
our bar ellipticity and fraction  with other studies in $\S$~4.3.

Another important result is that deprojection does not make any 
significant changes to the global bar fraction, when dealing with the 
fairly large OSUBSGS sample. As shown by Table 2 and Figure  8, 
the $B$- and  $H$-band bar fractions are 45\% and 58\% before 
deprojection, and change by only a factor of 0.97 and 1.03, 
respectively, after deprojection. 
We suggest several reasons for the small impact of 
projection effects.
First, this study uses only moderately inclined 
($i<$~60$^\circ$) galaxies where projection effects are 
less severe than in highly  inclined systems. 
Second, projection effects  produce large  changes in the 
morphology of a galaxy  only  when  the disk inclination, $i$,  
is significant {\it and}  the  difference in  PA  between  
the bar and the disk major axes is  
close to 90$^\circ$. From a statistical point of view, 
these two  conditions are unlikely to occur simultaneously in a 
dominant fraction of the  sample. 
These arguments are supported by Figures 9a and 9b, which  show that 
the galaxy classes  assigned  prior to deprojection are in no way 
biased by the galaxy  inclination, $i$: both barred and unbarred 
galaxies span a similar range in $i$. Furthermore, even  the 
bar ellipticity $e_{\rm bar}$ measured before deprojection is
uncorrelated with $i$ (Figs. 9c and 9d).

The fact that the bar fraction  in large samples is similar 
before and after deprojection is encouraging for large studies 
of bars at intermediate redshift  (e.g., Jogee \etal  2004, Elmegreen 
\etal 2004, Zheng \etal  2005), where 
deprojection is  not done because of the difficulty in accurately
measuring the PA of the line of nodes and the inclination of the
outer  disk in noisy images of distant galaxies.

\subsection{Sizes of bars and disks at $z\sim$~0}

As outlined in $\S$ 3.1, we use  the semi  major axis 
$a_{\rm bar}$, where the  bar ellipticity is a maximum,  
as a measure of the bar length.
We caution that this may underestimate the bar length  in some galaxies.
However, a  visual comparison of $a_{\rm bar}$ with the images of 
our galaxies suggests that $a_{\rm bar}$ does a reasonable job in most
cases.

The distributions of bar sizes or semi-major axes ($a_{\rm bar}$) 
before and after deprojection are shown  for the $B$ and $H$ 
bands in Figure 10.  
Some bars do appear larger after deprojection, but from a statistical 
point of view, deprojection does not have a substantial  effect on 
the bar size distribution. For example, the mean bar size in the $H$
band before deprojection is 3.4 kpc and after deprojection it is
4.0 kpc. 
Sizes of large-scale bars in the local Universe lie in the range $\sim$~1 to 14 
kpc, with most (68\% in $B$ and 76\% in $H$ ) bars 
having $a_{\rm bar} \le$ 5 kpc, and $\sim$~50\% of them clustering with 
 $a_{\rm bar}$  in the range  2 to 5 kpc.  
If such a distribution of bar sizes is present at a 
redshift $z \sim$~1, where $1\arcsec$ corresponds to 
8.0 kpc,
then only observations with angular resolutions  superior to 
$0\farcs3$  can  adequately resolve the majority of bars.
This is relevant for assessing the relative effectiveness of  
current NIR capabilities, such as  NICMOS,  and those of 
future planned missions, such as WFC3, in detecting high 
redshift bars  in the NIR band over wide fields.

In  Figure 11, we plot the bar size versus the disk size
before  and after deprojection. 
The  bar size is measured from the $H$ band,  whose  low 
extinction enables more accurate measurement than in the
optical. The disk is  measured from the $B$-band image,  which is
deeper than the $H$ band and traces the disk further out 
($\S$ 3.2). 
Both before and after deprojection, we find that  bar and disk sizes  
are correlated   with an average slope of $\sim$~0.9, albeit 
with a large scatter of several kpc in bar size at a given disk 
size. 

Figure 12 shows the observed bar semi-major axis distribution 
normalized to $R_{\rm 25}$ (the radius in arcseconds of the isophote, 
where the surface
brightness equals 25  mag arcsec$^{-2}$) of the disk. $R_{\rm 25}$
values are obtained from the Nearby Bright Galaxies Catalogue (Tully
1988; hereafter NBG), except for NGC 6753, 6782, 5078,
6907, 7814, and ESO 142-19, which are from the RC3.  
The ratio ($a_{\rm bar}$/$R_{\rm 25}$)  
lies primarily in the range 0.1 to 0.5 in 
both the  $H$ and $B$ bands (Fig. 12).  
Only a minority of galaxies have larger values out to 0.95.

These results are  consistent with several smaller earlier
studies. 
Laine \etal (2002) find that the sizes of primary bars  correlate with 
the host galaxy sizes and the ($a_{\rm bar}$/$R_{\rm 25}$) ratio 
lies  primarily in the range 0.1 to 0.5. 
Menendez-Delmestre et al. (2004)  find an average  
($a_{\rm bar}$/$R_{\rm 25}$) ratio of 0.35, on the basis 
of ellipse fits of 134 2MASS galaxies. 
In his  study of bar lengths, based on ellipse fits 
of  $R$-band images of  65 local  early-type S0-Sab galaxies, 
Erwin (2005)  finds a similar mean   
($a_{\rm bar}$/$R_{\rm 25}$) ratio of 0.38 and  reports 
a correlation between  bar size and disk size.

What do these results imply? From a  theoretical standpoint,  the 
size of the bar ($a_{\rm bar}$) depends on the concentration of matter
in the disk and the distribution of resonant material that can absorb
angular momentum from the bar (Athanassoula 2003). 
Furthermore, the prevalence of chaotic orbits between the  
4:1 and the  corotation resonance (CR)  would naturally 
lead bars to end somewhere between the two resonances.
If bars end very near the CR as is found observationally  
(e.g., Merrifield \& Kuijken 1995; Debattista \etal 2002;  
Aguerri \etal 2003), then our result that ($a_{\rm bar}$/$R_{\rm 25}$) 
is generally well below 1.0  suggests that the  CR of disk galaxies  
lies well inside their  $R_{\rm 25}$ radius. Furthermore, the 
correlation between bar and disk sizes and the narrow range in 
($a_{\rm bar}$/$R_{\rm 25}$)  suggests that the growths of the bar 
and disk may be intimately tied.

\subsection{Distribution of bar strengths as characterized by  $e_{\rm bar}$ at  $z\sim$~0}

The term `bar strength'  is not well defined in the literature.
Various measures of bar strength are used and each  
measure has some benefits and trade-offs.
These  measures include  the $Q_{\rm b}$ method (Block \etal 2002; 
Buta \etal 2003; Buta \etal 2005), the maximum 
ellipticity of the bar, bar/interbar contrasts,  Fourier
decomposition techniques (Elmegreen \& Elmegreen 1985; 
Elmegreen \etal 1996), and  visual estimates of strength (e.g.,  
Martin 1995; Eskridge \etal 2000, 2002) gauged via eyeball inspection 
of images.

The $Q_{\rm b}$ method (Block \etal 2002; Buta \etal 2003; 
Buta \etal 2005) directly  measures the 
gravitational torque exerted by the bar, but it measures the torque at 
only  one point along the bar. The $Q_{\rm b}$ method depends on 
the scale height of the disk and the ability to derive a reliable 
model for the potential using images. It is hard to apply this method 
to a large number of  intermediate redshift galaxies 
due to resolution and signal-to-noise limitations. 
In the bar/interbar
contrast method used by Elmegreen \& Elmegreen (1985) and Elmegreen
\etal (1996), the bar
strength is characterized by the ratio of the peak surface brightness
in the bar region to the minimum surface brightness in the interbar
region. The Fourier decomposition method also used by Elmegreen \&
Elmegreen (1985) and Elmegreen \etal (1996) is similar to the $Q_{\rm
  b}$ method. It characterizes bar strength by measuring the relative
amplitudes of the Fourier components of the bar. The maximum amplitude
of the m=2 mode determines the strength of a bar. 

In studies where ellipse fits are used to characterize bars, 
the maximum  ellipticity of the bar ($e_{\rm bar}$) is  used 
as a measure of bar strength (e.g., Athanassoula 1992a; Martin 
1995; Wozniak \etal 1995; 
Jogee \etal 1999, 2002a,b; Knapen \etal 2000; Laine \etal 2002). 
One advantage of this approach is that the bar ellipticity  
can be estimated without making any assumptions about  
the mass to light ratio of the  galaxy or its scale height. 
It can also be applied to local galaxies as well as galaxies
out to intermediate redshifts  ($z\sim$~0.2--1.0 ; Jogee \etal 
2004, Elmegreen \etal 2004). 
There are also several theoretical reasons that support the 
use of the maximum  bar ellipticity as a measure of bar strength.
Shen \& Sellwood (2004) compare bar strength in 
N-body simulations, as characterized by
the $m$~=~2 Fourier components and the peak ellipticity. 
They find  that the ellipticity is very well correlated to  bar strength
estimator $A$, where $A$ is the relative amplitude of the bisymmetric 
($m$~=~2) Fourier component of the mass density averaged over a certain
inner radial range where the bar dominates.
In addition, from an observational standpoint, Laurikainen \etal (2002) 
find that, on average, the gravitational torque, $Q_{\rm b}$,  and 
$e_{\rm bar}$  are correlated for $e_{\rm bar} \le$~0.6. 
For higher  $e_{\rm bar}$ values, the relation appears to flatten out  
although the small number of galaxies precludes a firm conclusion.

Nonetheless, if we  deem that a measure of bar strength 
should give an indication of the gas inflow rate that a bar
drives via gravitational torques, then  the maximum  ellipticity 
of the bar ($e_{\rm bar}$)   is only  a 
{\it partial} measure of the   bar strength.   Both the 
mass and shape of the bar influence the magnitude of the gravitational 
torque  at each point along the bar.  The  peak bar ellipticity
describes the shape of the bar, but does not directly measure  
its mass or luminosity.  While bearing  this caveat in mind, we 
use the maximum bar ellipticity  $e_{\rm bar}$ as a partial measure of 
the bar strength in this study.

Figure  13 shows the  observed and deprojected 
distributions of bar strength as characterized by $e_{\rm bar}$ 
from ellipse-fits in the $B$ (Figs. 13a,c)  and $H$ bands (Figs. 13b,d). 
It is striking  that only a very small  proportion 
(7\% in $B$; 10\% in $H$) of bars  are very weak with  
0.25~$ \le  e_{\rm bar} \le$~0.40, while 
the majority of bars  (70\% in $B$; 71\% in $H$)  
have moderate to high strengths as characterized by $e_{\rm bar}$, 
with  0.50~$ \le  e_{\rm bar} \le$~0.75.
This point is further illustrated in Figure 14, which
is a generalized plot of the fraction 
of disks with `strong' and `weak' bars.
It  shows how the fraction of spiral galaxies that  host bars
with  ellipticities ($e_{\rm bar} > e_{\rm 1}$) changes 
as we vary $e_{\rm 1}$. 
As we increase $e_{\rm 1}$  from 0.35 to 0.45, 0.55, and 0.75, 
the deprojected bar fraction  in the $B$ band falls  from 
43\% to  39\%, 34\%, and  7\%, respectively. 
Correspondingly, the  bar fraction  in the $H$ band falls from 
59\% to 47\%, 30\%, and  1\%,  respectively. 
The flattening of the curve around  $e_{\rm 1} \sim$~0.40 
shows that  the majority of bars have  $e_{\rm bar}$ above this 
value.
This has  implications for theoretical models 
that address the robustness of 
bars,  and we refer the reader to  $\S$~4.6 for a discussion.

How do our results on bar strength as characterized by 
the maximum  bar ellipticity $e_{\rm bar}$ from 
ellipse-fitting  compare with those of 
Buta \etal (2005) who use the $Q_{\rm  b}$ parameter? 
At first glance, the results may seem contradictory: they conclude that 
40\% of the galaxies in the OSUBSGS $H$ band have `weakly barred' 
or unbarred states ($Q_{\rm  b} \le 0.1$), whereas we find that 
only 6\%  of galaxies have `weak' bars with  
$0.25 \le e_{\rm bar} \le 0.4$  in the $H$ band after deprojection. 
However, it should be noted that Buta \etal (2005) 
group  unbarred and weakly barred galaxies together. 
Their cited fraction of  40\% for weak and unbarred states is, 
in fact, fully consistent with the fraction (46\%) that we find when 
we group together unbarred galaxies 
(40\%) and `weakly barred' galaxies (6\%).

How do the bar classes and bar strengths from ellipse-fits, as 
derived by our quantitative method  ($\S$~3.3), compare with the 
RC3  bar classes based on visual
inspection of optical $B$  images (de Vaucouleurs \etal 1991)?
The three  RC3 visual bar classes, `A', `AB', and `B' denote `unbarred', 
`weakly barred', and  `strongly barred' disks, respectively.
Of the 42, 47, and 46 galaxies in our sample that have an RC3 bar class of 
`A', `B', and `AB', respectively, our quantitative characterization 
($\S$~3.3) shows that 5\%, 85\%, and 41\% host bars in $B$-band images
and 19\%, 87\%, and 65\%  host bars in $H$-band images.
Clearly, only a small fraction  (41\% or 19/46)  of  galaxies with RC3 
bar class `AB'  qualify as barred in $B$-band images, according  to our
 quantitative criteria ($\S$~3.3).  We visually inspected the remaining 
27 galaxies that fail to qualify in order to investigate why they do not.
We  found that for 17 of them, we could  not identify a bar feature 
in the $B$-band image, even by eye. For the remaining 10, we could
visually see a somewhat elongated feature, but it does not satisfy 
the ellipticity and PA criteria outlined in $\S$~3.3. 
Another interesting point highlighted by  Figure 15  is that while 
the mean bar strength (as characterized by $e_{\rm  bar}$)  
is higher for RC3 visual class 
`B' than for class `AB',  the two classes have significant overlap in 
the range  $e_{\rm  bar} \sim$~0.5--0.7. Thus, RC3 bar types 
should  be used with caution and may be misleading.

It is also noteworthy that Figure 13 shows no evidence for  bimodality  
in the distribution of bar strength, 
as characterized by  $e_{\rm bar}$ from ellipse fits, 
in the $B$ or $H$ bands, in agreement 
with Buta \etal (2005). What about the bimodality claimed in earlier 
studies by  Abraham \& Merrifield (2000) and Whyte \etal (2002)? 
Both of these studies used the parameter $f_{\rm bar}$ to characterize 
the ellipticity  of the most elliptical feature of a galaxy, and  
measure  $f_{\rm bar}$ for both barred and unbarred galaxies.
They report no bimodality in $f_{\rm bar}$ among barred galaxies, which is 
consistent with our findings that  $e_{\rm bar}$ shows no  bimodality among 
barred galaxies. The only bimodality that they report in $f_{\rm bar}$ is 
between barred and unbarred galaxies. It is unclear how  robust this  
bimodality is since  Whyte \etal (2002) report  a bimodality  that is 
much weaker than the one seen by  Abraham \& Merrifield (2000).
The authors assigned this weakening to the larger sample size used
by Whyte \etal (2002). At any rate, we cannot make any direct comparison 
with  their bimodality results involving unbarred galaxies, since we measure 
$e_{\rm bar}$  in barred galaxies, but not in unbarred galaxies.   The  
reason for this selective measurement is rooted in our rigorous approach 
for identifying a bar. In the study of  Abraham \& Merrifield (2000) and 
Whyte \etal (2002),  a bar is simply considered as  the innermost feature 
whose isophote has the highest ellipticity.
In contrast, we use a rigorous approach for identifying a bar: we call
a feature a bar only if its radial variation of ellipticity and PA  follows 
the behavior expected based on the dominant orbits of a barred potential, 
as outlined in $\S$~3.3. We measure the maximum bar ellipticity $e_{\rm bar}$ 
only for those features that qualify as a bar.

\subsection{Bar fraction and ellipticity as a function of Hubble 
type at $z\sim$~0}

Figure 16 shows how the fraction of barred disks varies 
across different Hubble types in  sample S4. The Hubble types 
are taken from RC3 and the bins represent S0, Sa/Sab, Sb/Sbc, 
Sc/Sd, and Sd/Sm. 
We first note that the bar fraction in different RC3 Hubble
types  does not change significantly  after deprojection, whether 
in the $B$  (Fig. 16a vs. 16d)  or $H$ (Fig 16b vs. 16e) band images.
This is again encouraging for large studies of bars at intermediate 
redshift  (e.g., Jogee \etal  2004, Elmegreen \etal 2004, Zheng \etal  
2005), where deprojection is not done for the reasons outlined 
in $\S$~4.1.

In  the $B$ band, we find that the bar fraction is lower with 
respect to the $H$ band by  ~$\sim$~1.2--1.5 in Sas to Scs, and by 
$\sim$~2.5 in Sds/Sms (Fig. 16c,f). This is  consistent with  higher 
obscuration in  dusty, gas-rich late types. Eskridge et al. (2000) 
also find that the increase in  bar fraction  from the $B$ to $H$ 
band is most significant for late-type  galaxies. 

How does the bar fraction vary across RC3 Hubble types? The number of 
galaxies involved are too small in the S0 and Sd/Sm bins for  
robust number statistics and we therefore restrict our analysis to 
types Sa to Scd. We conclude that the  $H$-band bar fraction (Fig 16e) 
remains  $\sim$~60\%  across RC3 Hubble types Sa to Scd.
Our quantitative result based on 136 galaxies is 
consistent with the results based on ellipse fits of 
a much smaller sample (58 galaxies) by Knapen, Shlosman,\& Peletier 
(2000), as well as with the qualitative results of Eskridge 
et al. (2000), who also  report  a constant NIR bar fraction 
as a function of RC3 Hubble types, based on  visual inspection.
The large  $H$-band bar fraction  of $\sim$ 60\%  across
different Hubble types  implies that bars are ubiquitous in  
spirals across the entire Hubble sequence. Further implications are 
discussed  in  $\S$~4.6.

How does the bar strength, as characterized by  $e_{\rm bar}$ 
from ellipse-fitting, vary as a function of RC3 Hubble type? 
In the $H$ band,  the  bar strength  $e_{\rm bar}$  lies in 
the range 0.35--0.80, and shows no systematic variation across 
Hubble types  Sa to Scd, either before (Fig. 17a) or after (Fig. 17b) 
deprojection.
We note, however, that Buta et al. (2004) and Laurikainen et al. 
(2004) find that the $Q_{\rm b}$ and  $Q_{\rm g}$ parameters tend to 
have lower values toward earlier-type galaxies. In order to 
understand this discrepancy, we first note that the $Q_{\rm b}$ and  
$Q_{\rm g}$ parameters  measure the bar strength 
relative to the axisymmetric components, such as the disk and bulge.
The lower  $Q_{\rm b}$ and  $Q_{\rm g}$ values in early type galaxies could 
reflect the fact that such galaxies  have stronger axisymmetric components, 
which make the relative strength of the bar lower, even if the bar was 
as strong or stronger intrinsically than those in later-type galaxies.

\subsection{Comparison of optical properties of bars 
at $z \sim$~0  and at $z \sim$~0.2--1.0}

Studies of bars  at $z \sim$~0.2--1.0  (lookback times of  3--8 Gyr) 
based on  $HST$ ACS observations in the  Tadpole field (Elmegreen \etal 2004), 
the GEMS and GOODS fields (Jogee \etal 2004), and COSMOS surveys (Sheth \etal in 
preparation) trace 
bars in the {\it rest-frame optical}.
The reddest ACS filter F850LP has a  pivot wavelength of 9103 \AA,  
while the value for the F814W  filter is 8064 \AA.  Over the redshift 
range  $z\sim$~0.2--1.0, the rest-frame wavelength traced by the F850LP 
filter ranges from  7586 \AA~to 4550 \AA, which corresponds to the {\it 
rest-frame optical} $R/I$ to $V/B$ bands.  
In order to avoid the pernicious effects of bandpass shifting, it is essential
that ACS studies of bars at $z \sim$~0.2--1.0  compare their rest-frame
optical results to  the optical bar fraction at $z \sim$~0,  rather 
than to the NIR bar fraction at  $z \sim$~0. If the NIR  $z \sim$~0 point is used 
for comparison (e.g, Menendez-Delmestre 
\etal 2006), it will lead to flawed  conclusions because the NIR $z \sim$~0 
bar fraction  (60\%~$\pm$~6\%) is significantly larger than the 
optical $z \sim$~0 bar fraction (44\%~$\pm$~6\%), as reported  in $\S$~4.1.
We therefore use the OSUBSGS optical bar fraction at $z \sim$~0  in the 
discussion below.

In the  study of bars  at $z \sim$~0.2--1.0, Jogee 
\etal (2004)  ellipse fitted a  sample of 1590 galaxies at $z \sim$~0.2--1.0, 
drawn from 25\% of the GEMS survey area.  
Then  they  applied essential cutoffs in absolute  magnitude, 
bar size, and  bar ellipticity  in order to ensure a complete sample, 
high spatial resolution, and reliable bar identification out to  $z \sim$~1.
In particular, in order to ensure that the sample 
of spiral galaxies is fairly  complete out to $z \sim$~0.9,  
an absolute magnitude cutoff of $M_{\rm V}<$~$-$19.3 had to 
be applied. 
Secondly,  at $z >$~0.5 (where  $1\arcsec$ corresponds to scales $>$~6.2 
kpc), the study  could not efficiently resolve very small bars 
with  semi-major axes   $a<$~1.5 kpc, in agreement with Lisker 
\etal (2006). Thus, a cutoff of  $a_{\rm bar} \ge$ 1.5 kpc is implicitly
applied.
Finally, the study only considered  bars with  moderate ellipticity  
$e_{\rm bar} \ge$~0.4  because at  intermediate redshifts, it becomes 
difficult to  unambiguously  identify and characterize bars with lower 
ellipticities. This is not a dramatic cutoff as most bars 
have $e_{\rm bar} \ge$~0.4 (Fig. 13). 
After applying these cutoffs in absolute magnitude 
($M_{\rm V}<$-19.3), bar size ($a_{\rm bar} \ge$ 1.5 kpc), and 
bar ellipticity ($e_{\rm bar} \ge$~0.4), Jogee \etal  (2004)  find  
a rest-frame  optical bar fraction of  $f_{\rm optical2}$$\sim$~30\% 
$\pm$ 6\%  $z \sim$~0.2--1.0.  
A constant and similar optical bar fraction  (23\% to 40\%) 
out to  $z \sim$~1 is also reported by  Elmegreen \etal (2004).

In order to get a  valid optical bar fraction for 
comparison at $z\sim$~0, we must apply the exact same cutoffs to 
the OSUBSGS optical data.
We start with  observed bar properties  prior to deprojection 
from OSUBSGS because  no deprojection  
was applied in any of the intermediate redshift studies 
(Jogee \etal  2004; Elmegreen \etal 2004; Zheng \etal  2005).
With a cutoff of $M_{\rm V}<$-19.3, the optical $B$-band bar fraction  
at $z\sim$~0 drops from 45\% (61/136) to 43\% (45/104).
Applying a further cutoff of $a_{\rm bar} \ge$ 1.5 kpc makes
it drop  to 36\% (37/104).
Finally, a third cutoff  of $e_{\rm bar} \ge$~0.4 reduces 
the  optical $B$-band bar fraction to  34\% (35/104).

Thus, after the same cutoffs in absolute magnitude 
($M_{\rm V}<$-19.3), bar size ($a_{\rm bar} \ge$ 1.5 kpc), and 
bar ellipticity ($e_{\rm bar} \ge$~0.4) are applied, 
a very good agreement ensues  between the  GEMS optical bar fraction  
at $z \sim$~0.2--1.0  ($f_{\rm optical2}$$\sim$~30\% $\pm$ 6\%) 
and  the OSUBSGS   optical $B$-band bar fraction at $z\sim$~0 
($f_{\rm optical3}$$\sim$~34\%~$\pm$~6\%). This agreement   
strongly suggests that the optical  bar fraction 
in bright disks does not decline strongly with redshift. 
Such a decline would cause  $f_{\rm optical2} \ll$~$f_{\rm optical3}$  
because the observed bar fraction would be lowered both 
by the intrinsic decline, and  by systematic effects at 
intermediate redshifts, such as cosmological dimming, 
the loss of spatial resolution, and lower signal-to-noise.

However, our finding allows for models 
where the  optical bar fraction is  
either constant, or  rises with redshift. 
In the latter class of models, one can arrive at comparable 
values of $f_{\rm optical2}$ and $f_{\rm optical3}$  only if 
the intrinsic increase in bar fraction with redshift produced
by the model  is compensated for  by the `loss' of bars due to 
systematic effects, such as cosmological dimming, and 
low signal-to-noise.  
In a forthcoming paper, we will assess 
the impact of such redshift-dependent  systematic effects 
by artificially redshifting the OSUBSGS sample 
to $z \sim$~1, and repeating the bar characterizations.
This will enable us to distinguish between the  two 
classes of models.

\subsection{Constraints on the robustness and evolution of bars}

The  robustness and lifetime of bars define some 
of the most fundamental issues in  the evolution of bars, 
their impact on  disk galaxies ($\S$~1) and  the assembly 
of the Hubble sequence. 
In general terms, the evolution of a bar depends 
on the exchange   of  angular momentum  
between  the stars in the bar and the other components 
of a galaxy, namely, the  dark matter (DM) halo and the baryons
(gas and stars)  in the bulge and disk.
Important factors influencing the bar include 
the triaxiality of the  DM  halo in 
which it lies (e.g., Berentzen, Shlosman, \& Jogee 2006); 
the amount of angular momentum that the DM  halo can  absorb 
(Athanassoula 2003);  
the central mass concentrations (CMCs) present  in the inner
few hundred pc (e.g.,  Shen \& Sellwood 2004;  
Athanassoula \etal 2005; Martinez-Valpuesta \etal 2006; 
Debattista \etal 2006); 
and the distribution and amount of gas in the disk 
(e.g, Shlosman \& Noguchi 1993; Bournaud et al. 2002, 2005; 
Debattista \etal 2006).  
In this section, we compare our empirical results to different 
simulations in order to constrain  theoretical scenarios.
We note, however, that most simulations do not yet fully 
incorporate the effects of star formation and feedback, 
which can impact the evolution of the disk in important
ways.

Dubinski (1994) showed that the triaxiality of DM
halos is diluted by baryonic dissipation.
Recent simulations by Berentzen, Shlosman, \& Jogee (2006)  find that
bars embedded in triaxial non-rotating DM halos can only survive
if the inner halo ellipticity is washed out.
Otherwise, the interaction between the bar and the DM halo
induces chaotic orbits and destroys the bar.
In the present paper, our findings that the majority (60\%) of
spirals are barred in
the infrared  ($\S$~4.1),  and that these bars have primarily
moderate to high strengths, as characterized by the maximum  
bar ellipticity $e_{\rm bar}$ (0.50~$ \le  e_{\rm bar} \le$~0.80; 
$\S$~4.3), suggest that DM halos of most present-day spirals are
close to axisymmetric, with a  maximum equatorial axial ratio of
$\sim$~0.9 in potential.
These limits may change slightly if one allows the DM halo to
have a figure of rotation.
These results are consistent with Kazantzidis \etal (2004), who
find that in the very early stages of disk formation,  the
settling of the dissipative baryonic component within a
triaxial halo strongly dilutes the triaxiality to such values.
Berentzen \& Shlosman (2006) also report that a growing disk is
responsible for washing out the halo prolateness (in the disk plane)
and for diluting its flatness over a period of time comparable to
the disk growth.

The CMC typically refers to the mass present within the inner 
hundred or few hundred pc. A large or more centrally concentrated CMC  
can weaken a bar amplitude by changing the orbital 
structure of a barred potential and inducing chaotic orbits.
Most recent simulations (e.g.,  Athanassoula \etal 2005; 
Shen \& Sellwood 2004; Martinez-Valpuesta \etal 2006; Debattista
\etal 2006) find that bars are more robust than previously thought: 
in order to produce any significant reduction in  bar strength, the  ratio 
$X_{\rm CMC}$~$\sim$~($M_{\rm CMC}$/$M_{\rm disk}$), 
where $M_{\rm CMC}$ is the mass 
of the CMC in the inner few hundred pc, and $M_{\rm disk}$ is the 
disk mass,  must be very  large, at least  10\%.  Such large 
values are only of academic interest and are not realized 
in present-day galaxies, as we discuss below.

In present-day galaxies,  the components that contribute to the
CMC in the inner few hundred pc consist of  super-massive 
black holes (SMBHs)  central dense stellar clusters,  gaseous concentrations, 
and the inner parts of bulges. SMBHs have typical masses in the range 
10$^{6}$--10$^{9}$~$M_\odot$ and tend to scale as 0.001
of the bulge mass;  gaseous concentrations 
range from  10$^{7}$--10$^{9}$~$M_\odot$ in the central
500 to 1000 pc radius (e.g., Jogee \etal 2005); and central dense 
stellar clusters typically  have masses in the range  
10$^{6}$--10$^{8}$~$M_\odot$. 
These components typically lead to  $X_{\rm CMC}$  values 
that are much lower than 10\%. 
This  suggests that CMCs that exist in present-day galaxies 
are not large enough to produce any significant reduction in  bar  strength.
Our results are consistent with these expectations and with
simulations  that support robust bars. 
We  found that  the majority  ($\sim$~71\%--80\%)  of bars have  moderate 
to high strengths, as characterized by  $e_{\rm bar}$ from ellipse-fitting  
(0.50~$ \le  e_{\rm bar} \le$~0.80).
We also found that the  bar fraction ($\sim$~60\%)  and mean 
bar strength, as characterized by  ellipse fits 
($e_{\rm bar}\sim$~0.5), is relatively 
constant across RC3 Hubble types Sa to Scd ($\S$~4.4), although 
the latter encompasses a wide range of  gas mass fractions, 
CMC masses, and CMC components.

Gas can affect the formation and evolution of a bar in 
different ways, depending on its distribution and clumpiness.
In the case of an  unbarred disk,  the  accretion of  
cold gas  makes the disk more massive, 
dynamically colder, and therefore more bar unstable 
(e.g., Bournaud \etal 2002).  However, in the case of  very gas-rich 
disks,  the gas can become clumpy,  and the effect of dynamical 
friction on massive gas clumps at low radii can heat the disk 
and prevent it from forming the bar (e.g., Shlosman \& Noguchi 1993).
In the case of a disk that is already barred, the bar exerts gravitational
torques that  drive gas  located outside the corotation resonance 
(CR) outward, and drive gas located between the CR and  inner Lindblad 
resonance (ILR) inward. 
Most simulations to date  (e.g.,  Debattista \etal 2006;  Berentzen 
\& Shlosman in preparation; Curir \etal 2006) suggest that 
gas inflows in  present-day galaxies do not readily destroy bars.
For instance,  simulations  (e.g., Debattista \etal 2006), 
can only destroy the bar  when there are large gas inflows that
build  a very massive, soft CMC,  of order 20\% of the mass of 
the total baryonic (gas and stars) disk.  
Furthermore, the simulations also suggest that gas which 
sinks into the center can become bar supporting if it forms stars.
As discussed above, CMCs as large as 10\% or 20\%  are not 
realized in present-day galaxies and the simulations therefore imply that 
gas inflows in  present-day galaxies do not readily destroy bars.
In the very early Universe, if extreme gas inflows and extreme CMCs are 
realized, the evolution of bars might be different.

We note that simulations of  bar-driven gas inflow by 
Bournaud \etal  (2005)  yield widely different predictions from 
those discussed above. 
The simulations of Bournaud \etal  (2005) appear to destroy a bar even with a  
gas  mass fraction (GMF) that is as low as 5\% to 7\%. Here, 
the GMF is defined as the ratio of gas mass to the total mass of 
the  stellar disk. A GMF of order 5\% is easily met in present-day 
galaxies and  these simulations would suggest, therefore, 
that strong bars in present-day galaxies are easily  destroyed 
by bar-driven gas inflows (Bournaud \etal  2005). There is clearly
a stark difference between the predictions of these simulations and 
the ones outlined in the previous paragraph. 
Part of the reason why the simulations yield such different results  
might lie in the  way the DM halo is modeled   
and the assumed ratio of  DM halo mass to disk mass.  
The DM halo is live and dominates over the disk mass 
in  Debattista \etal (2006), while it is rigid and 
less massive than the disk in Bournaud \etal  (2005).

What do our observational results suggest?  
We found that at $z\sim$~0, only a small  fraction ($\sim$~7\%--10\%)  
of bars are very weak (0.25~$ \le  e_{\rm bar} \le$~0.40), while 
the majority  ($\sim$~71\%--80\%)  of bars have moderate to high 
strengths  (as characterized by the maximum  bar ellipticity $e_{\rm bar}$),
with 0.50~$ \le  e_{\rm bar}\le$~0.80.  We also do not see any sign of 
bimodality in bar strength, as characterized by 
$e_{\rm bar}$ from ellipse fits.  Finally, we found that the bar fraction 
($\sim$ 60\%)  and mean  bar ellipticity  ($e_{\rm bar}\sim$~0.5) 
is relatively constant  across RC3 Hubble types Sa to Scd ($\S$~4.4),  
despite the wide variation  in GMFs. 
Our results are easily reconciled with scenarios 
where bars  in  present-day  moderately gas-rich galaxies 
remain strong under the effect of  bar-driven gas inflows.
Our results do not  necessarily rule out models  
where bars are easily destroyed by bar-driven gas inflows.
They do, however, imply that  if such an easy destruction 
occurs, then  there  must be  a very efficient  mechanism 
that not only regenerates  bars on a short timescale 
(e.g., Block \etal 2002; Bournaud \etal 2002),  but is 
also very well tuned to the bar 
destruction rate so that it can reproduce the observed  
constant  optical bar fraction in bright galaxies over the last 8 Gyr 
(Jogee \etal 2004;  Elmegreen \etal 2004;~$\S$~4.5).


\section{Summary and Conclusions}
 With the advent of high redshift $HST$ surveys, such as 
the Tadpole Field, GEMS, GOODS, and COSMOS, which  
trace bars in the rest-frame optical band out to $z\sim$~1,
it becomes increasingly important to provide a reference baseline
for bars at $z\sim$~0 in the optical band.
Motivated by these considerations, we characterize the frequency and 
structural properties of bars  at $z\sim$~0 in the  optical and NIR 
bands, by   ellipse-fitting the $B$ and $H$ images of 180 spirals in 
the OSUBSGS (Eskridge \etal 2002), and applying quantitative criteria 
in order to identify and characterize bars.  
We determine the inclination of the outer disk and exclude highly 
inclined  ($i >$~60$^\circ$) galaxies to derive a sample S4 of
136 moderately inclined  spirals. For this sample, we derive bar 
properties  both before and after deprojection to face-on. Our study 
complements existing work on OSUBSGS based on Fourier amplitudes 
(Block \etal 2002; Buta \etal 2005) and visual classification 
(Eskridge \etal  2000), and it can be compared with studies 
(Jogee \etal  2004; Elmegreen \etal 2004; Zheng \etal  2005) of 
intermediate redshift ($z \sim$~0.2--1.0) bars employing the same 
ellipse-fitting methodology. Our results are summarized below.

{\it \rm
(1)~The optical and NIR bar fraction at $z\sim$~0:
}
For our sample, which  is dominated by galaxies with 
$M_{\rm V} \sim$~-20 to -22, we find a deprojected  bar 
fraction at $z\sim$~0 of 
$f_{\rm NIR1}$~$\sim$~60\%~$\pm$~6\% in the near-infrared $H$ band, and 
$f_{\rm optical1}$$\sim$~44\%~$\pm$~6\%  in the optical $B$-band images.
The latter  likely miss  bars obscured by dust and star formation.
Deprojection does not make any significant changes 
to the global  $B$- and  $H$- band bar fractions, which are 
 45\% and 58\% before  deprojection, and change by only a factor 
of 0.97 and 1.03, respectively, after deprojection. 
This is encouraging for large studies of bars at intermediate redshift  
(e.g., Jogee \etal  2004, Elmegreen \etal 2004, Zheng \etal  2005), 
where deprojection is not performed.

{\it   \rm
(2)
Comparison of optical properties of bars at $z \sim$~0  and at  intermediate redshifts:
}
Studies of bars  at $z \sim$~0.2--1.0  (lookback times of  3--8 Gyr) 
based on  $HST$ ACS observations in the  Tadpole field, the GEMS and 
GOODS fields, and COSMOS surveys trace bars in the {\it rest-frame optical}.
$R/I$ to $V/B$ bands (7586 \AA~to 4550 \AA).  
Therefore, in order to avoid the pernicious effects of bandpass shifting, it is 
essential that ACS studies of bars at $z \sim$~0.2--1.0  compare their rest-frame
optical results to the optical bar fraction at $z \sim$~0,  rather than to 
the significantly higher  NIR bar fraction at  $z \sim$~0. 

\noindent 
Furthermore, at $z \sim$~0.2--1.0, it is 
essential to apply  cutoffs  in absolute magnitude,
bar size, and  bar ellipticity in order to ensure a complete sample, 
adequate spatial resolution, and reliable bar identification. 
After applying  cutoffs in absolute 
magnitude ($M_{\rm V}<$-19.3), bar size ($a_{\rm bar} \ge$ 1.5 kpc), and 
bar ellipticity ($e_{\rm bar} \ge$~0.4),  
Jogee \etal (2004) found a  rest-frame  optical bar fraction 
of  $f_{\rm optical2}$$\sim$~30\%~$\pm$ 6\% at  $z \sim$~0.2--1.0.  
A constant and similar optical bar fraction  (23\% to 40\%) 
out to  $z \sim$~1 is also reported by  Elmegreen \etal (2004).
In order to derive the equivalent optical bar fraction for 
comparison at $z\sim$~0, we applied the exact same cutoffs to 
the OSUBSGS optical data. 
With a cut off of $M_{\rm V}<$-19.3, the optical bar fraction  $z\sim$~0 
drops from 45\% (61/136) to 43\%.
Applying a further cutoff of $a_{\rm bar} \ge$ 1.5 kpc makes
it drop  to 36\%. 
Finally, a third cutoff  of $e_{\rm bar} \ge$~0.4 reduces 
 optical $B$-band bar fraction at $z\sim$~0  to
$f_{\rm optical3}$$\sim$~34\%~$\pm$~6\%.
The result that  $f_{\rm optical2}$ is comparable to $f_{\rm optical3}$
rules out scenarios where the {\it optical}  bar fraction in 
{\it bright} disks  declines strongly with redshift. It allows for 
models where the  optical bar fraction is 
either constant, or  rises with redshift.

{\it  \rm
(3) Distribution of bar strengths  $z\sim$~0  as characterized by  
ellipse-fitting:}
In this study, we use the maximum  bar ellipticity   $e_{\rm bar}$ 
from ellipse-fits as a partial measure of the bar strength.
Only a  very small proportion 
(7\% in $B$; 10\% in $H$) of bars  are very weak  as
characterized by  $e_{\rm bar}$ from ellipse fits 
(0.25~$ \le  e_{\rm bar} \le$~0.40), while 
the majority of bars  (70\% in $B$; 71\% in $H$)  
have moderate to high ellipticities  (0.50~$ \le  e_{\rm bar} \le$~0.75). 
We find  no evidence for  bimodality  
in the distribution of bar strength, as characterized by  $e_{\rm bar}$ 
in the $B$ or $H$ bands, 
in agreement with Buta \etal (2005).

{\it  \rm
(4)
Bar fraction and strength, as characterized by  ellipse-fitting, 
as a function of RC3 Hubble type at $z\sim$~0:}
The  deprojected bar fraction is 60\% in $H$ and  44\% in $B$, 
confirming the ubiquity of local bars. 
In  the $B$ band, the bar fraction is lower with 
respect to the $H$ band by  ~$\sim$~1.2--1.5  for Hubble types 
S0s to Scs, and by $\sim$~2.5  for Sds/Sms. This is consistent 
with the higher obscuration in  dusty, gas-rich late types.
The bar fraction and bar strength,  as characterized by  $e_{\rm bar}$, 
in the  $H$ band  shows no systematic variation across Hubble types 
Sa to Scd.

{\it  \rm
(5) Comparison with RC3 visual bar classes:
}
Of the 42, 47, and 46 galaxies in our sample that have an RC3 visual 
bar class of  `A' (unbarred), `B' (strongly barred), and `AB' (weakly 
barred), respectively, our quantitative characterization 
($\S$~3.3) shows that 5\%, 85\%, and 41\% host bars in $B$-band images
and 19\%, 87\%, and 65\%  host bars in $H$-band images.
Thus, quantitative characterization of bars differs significantly
from RC3 bar classes for the RC3 bar class `AB'.
Furthermore, the mean bar strength, as characterized by 
the maximum  bar ellipticity $e_{\rm  bar}$,  is higher 
for RC3 visual class `B' than for class `AB', but 
the two classes have significant overlap in the range 
$e_{\rm  bar} \sim$~0.5--0.7. Thus, RC3 bar types should 
be used with caution and may be misleading.

{\it  \rm   
(6)~Sizes of bars and disks at $z\sim$~0: 
}
The sizes or semi-major axes  $a_{\rm bar}$  of large-scale bars 
in the local Universe lie in the range $\sim$~1 to 14 
kpc, with  the majority  of bars (68\% in $B$ and 76\% in $H$) 
having  $a_{\rm bar} \le$~5 kpc. Bar and disk sizes  
are correlated   with an average slope of $\sim$~0.9, albeit 
with a large scatter of several kpc in bar size at a given disk 
size.  The ratio ($a_{\rm bar}$/$R_{\rm 25}$) lies 
primarily in the range 0.1 to 0.5, with only a minority of 
galaxies having larger values out to 0.95.
The correlation  between bar and disk sizes, and the narrow range 
in $a_{\rm bar}$/$R_{\rm 25}$  
suggests that the growths of the bar and disk may be intimately
tied. The fact that ($a_{\rm bar}$/$R_{\rm 25}$) is generally well 
below 1.0  suggests that the  CR of disk galaxies  lies well 
inside their  $R_{\rm 25}$ radius, assuming that  bars end  
near the CR.

{\it   \rm
(7)
Constraints on the robustness of bars:
}
Our findings that the majority (60\%) of spirals are barred in 
the infrared  and that most  ($\sim$~71\%--80\%) of  these bars have primarily 
moderate to high ellipticities (0.50~$ \le  e_{\rm bar} \le$~0.80)  
suggest that DM halos of present-day spirals 
have at most a mild triaxiality, with a maximum equatorial 
axis ratio $b$/$a$~$\sim$~0.9 in  the potential. 
We also found that the bar fraction   
and mean bar strength  (as characterized by 
the maximum  bar ellipticity $e_{\rm  bar}$)
are relatively constant  across Hubble types Sa to Scd, 
and there is no  bimodality in $e_{\rm  bar}$.
Taken together, our results are easily reconciled with scenarios 
where bars  in  present-day  galaxies are  relatively
robust against the range in  gas mass fractions, gas inflows, 
and CMC components  present across Hubble types Sa to Scd.
 Our results do not  necessarily rule out models  
where bars are easily destroyed by bar-driven gas inflows.
They do, however, imply that  if such an easy destruction 
occurs, then  there  must be  a very efficient  mechanism 
that not only regenerates  bars on a short timescale,  but is 
also very well tuned to the bar destruction rate so that it can 
reproduce the observed  constant  optical bar fraction in bright 
galaxies over the last 8 Gyr.

S.J. and I.M. acknowledge support from NSF grant AST-0607748, 
NASA LTSA grant NAG5-13063, as well as 
$HST$ grants G0-1048  and G0-10395 from STScI, which is 
operated by AURA, Inc., for NASA, under NAS5-26555. 
The Ohio State University Bright Spiral Galaxy Survey, was funded by
grants AST-9217716 and AST-9617006 from the United States National Science
Foundation, with additional support from the Ohio State University. 
We thank Pat Shopbell, Peter Teuben, and Stuart Vogel for their 
 assistance with the  Zodiac and MIRIAD packages; 
Seppo Laine for sharing his deprojection code from Laine 
et al. (2002); and James Davies for help with IRAF and IGI 
visualization  routines.
We also thank Paul Eskridge, Ron Buta, Fabio Barazza, Debbie Elmegreen, 
Isaac Shlosman, Ingo  Berentzen, Seppo Laine, Juntai Shen, Victor Debattista, 
Lia Athanassoula, Francoise Combes, Frederic Bournaud, Jerry Sellwood, 
and Johan Knapen for useful discussions.




{}
                                                                                


%

\clearpage
\setcounter{figure}{0}
\begin{figure}
\epsscale{.70}
\plotone{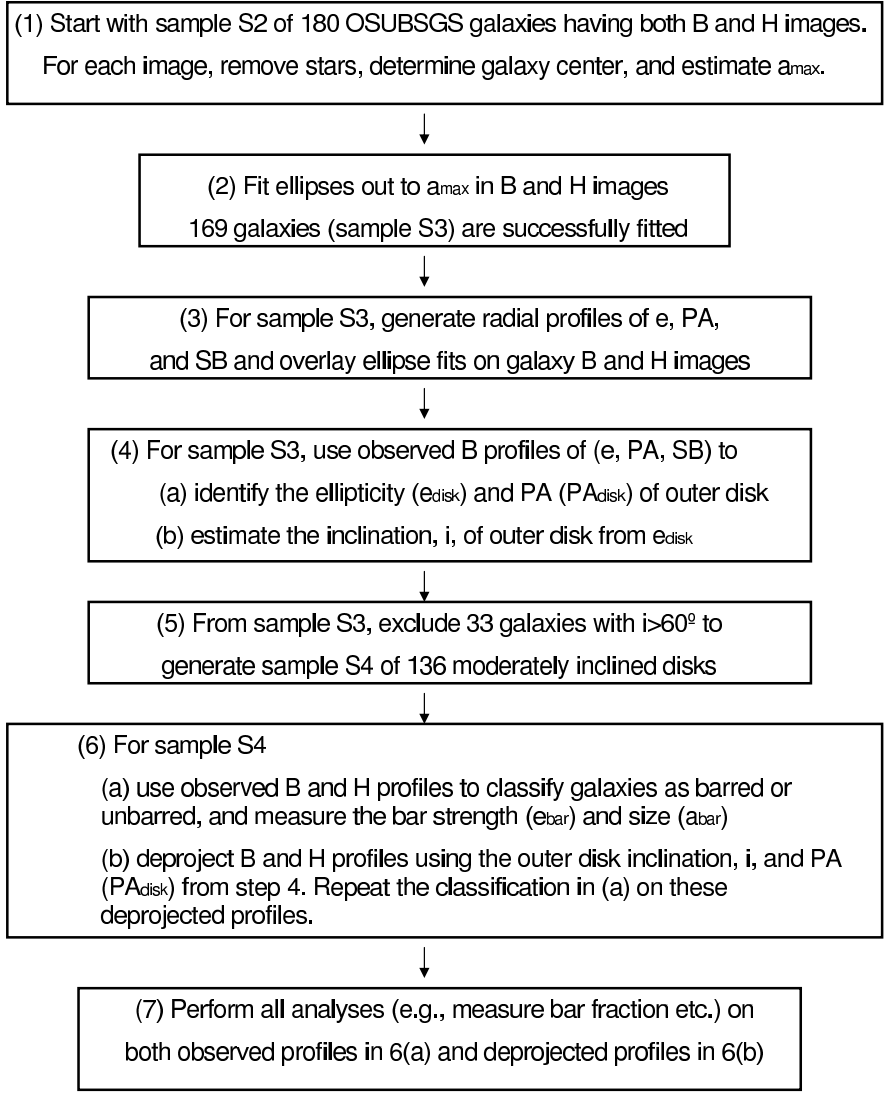}
\caption{
\bf 
Analysis steps for characterizing bars and disks  at $z\sim$~0 from OSUBSGS:
\rm
\noindent
Our procedure of characterizing bars and disks in OSUBSGS galaxies 
via ellipse fits is schematically illustrated in this figure and
described in  detail in $\S$~3.1--3.4.
For the $B$ and $H$-band images of the 180 
galaxies in sample $S$2, we  remove stars,  find an accurate center, 
and determine the maximum semi-major axis of the galaxy, $a_{\rm max}$,
where the galaxy isophotes reach the sky level. 
We fit ellipses out to $a_{\rm max}$ to  the $B$ and $H$ images 
of each galaxy, generate radial profiles of $e$, PA, and SB, and 
overlay the ellipses on the galaxy image for inspection. 
Successful fits are found in both bands for 169 galaxies (sample S3). 
For sample S3, we use the $B$-band radial profiles to characterize 
the inclination $i$ and PA of the outer disk. We exclude 33 galaxies 
with $i > 60^\circ$ to generate sample S4 of 136 moderately 
inclined galaxies. For sample S4, we deproject the $B$ and $H$ radial 
profiles using the outer disk $i$ and PA, and use the deprojected profiles to 
characterize the properties of barred and unbarred disks. 
For completeness, we also perform this characterization on the 
the observed profiles before deprojection.
} 
\end{figure}

\clearpage\setcounter{figure}{1}
\begin{figure}
\epsscale{0.8}
\plotone{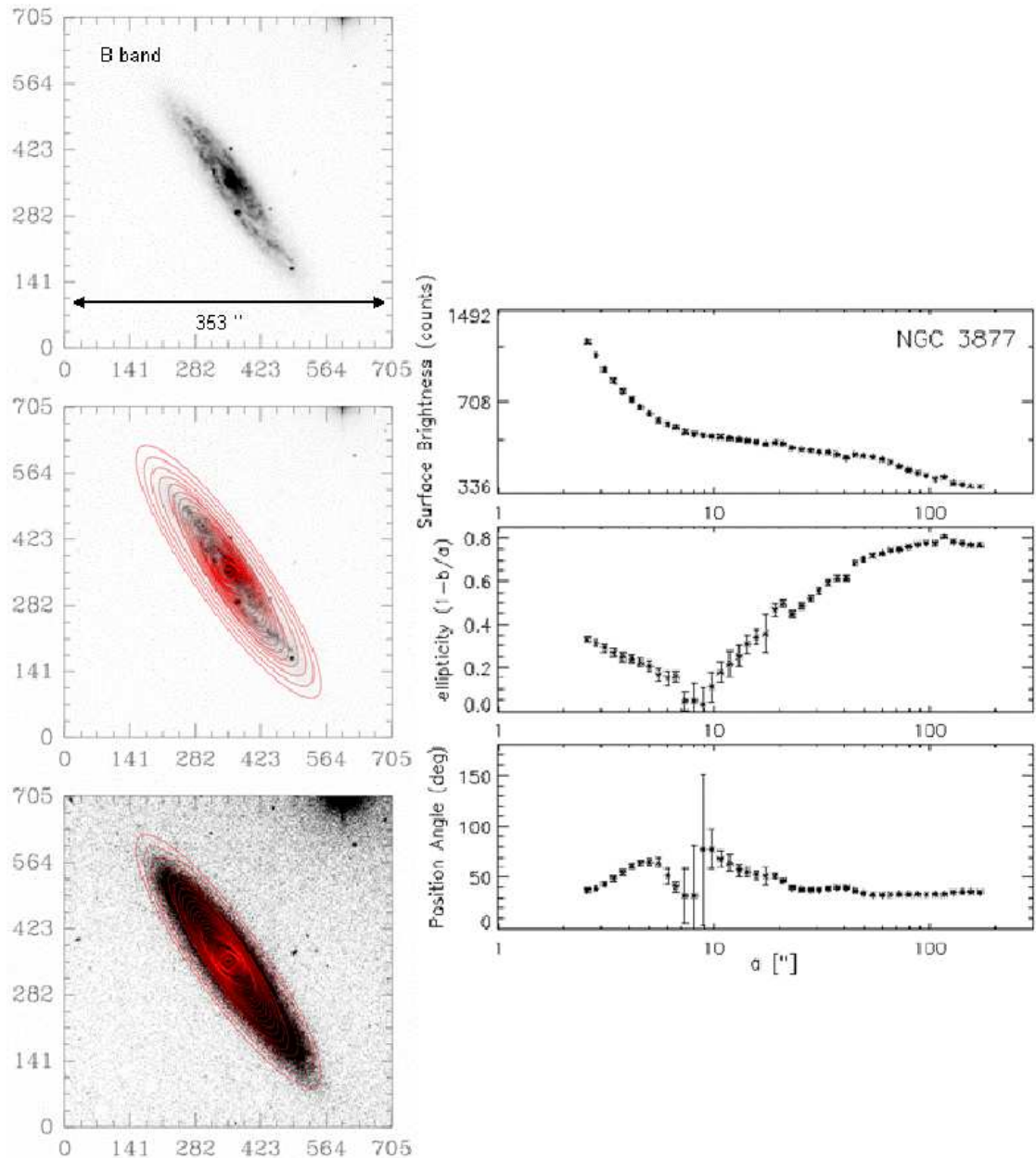}
\caption{\bf 
Ellipse fits to the $B$-band image of an inclined ($i > 60^\circ$) galaxy:
\rm
The left panel made of 3 images shows the ellipses fitted to the $B$-band 
image of NGC 3877. The top image shows only the galaxy. The scale is shown on the
 top image in arcseconds, where 1$\arcsec$ = 86 pc.
The middle and bottom images show 
the  ellipses overlaid on the galaxy, with  greyscale stretches chosen to 
emphasize the inner  (middle image) and outer (bottom image) regions of the 
galaxy. Note that ellipses are fitted out to the sky level in the image.
The right panel shows the radial profiles of surface brightness (SB), 
ellipticity ($e$), and position angle (PA)  versus semi-major axis ($a$) 
derived from the ellipse fits.
The profiles show evidence for some structure in the inner regions, but at 
$a >$ 100$\arcsec$, the $e$ settles  to a high value of 0.8, while the PA 
also settles to a constant value. This is the signature of an inclined 
disk with $i > 60^\circ$.
} 
\end{figure}

\clearpage
\begin{figure}
\setcounter{figure}{2}
\plotone{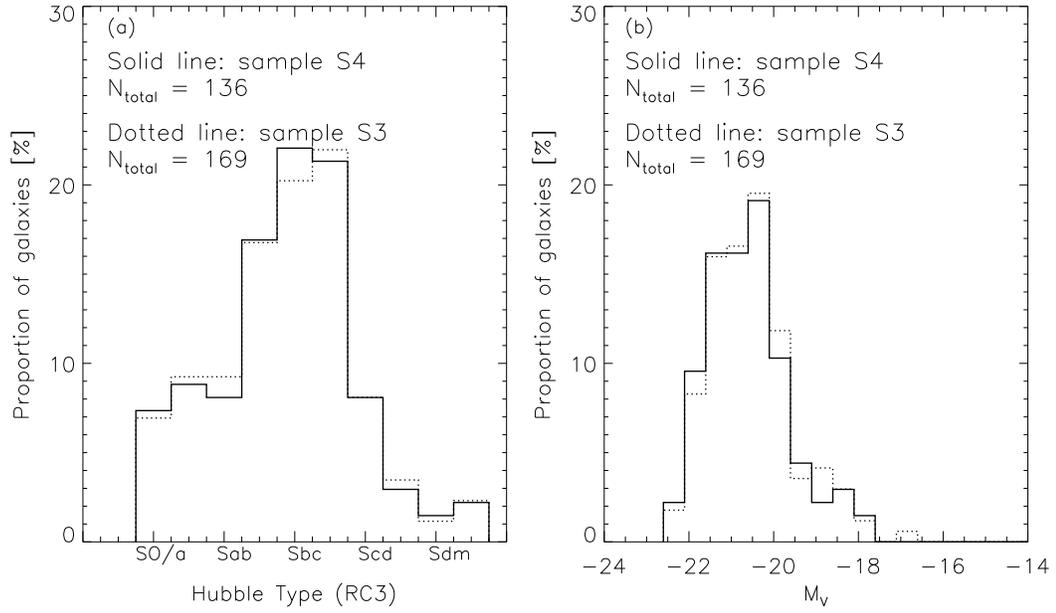}
\caption{
\bf 
Distribution of  Hubble types and absolute magnitudes:
\rm 
Left: The  distributions of  RC3 Hubble types are shown  
for the sample S4 (solid line) of 169 galaxies that include 
inclined systems, and for the sample S3 (dotted line) 
produced by excluding 33 galaxies with high inclination 
($i > 60^\circ$). This exclusion does 
not significantly affect the Hubble type distribution of the sample. 
Right: The distributions of  absolute $V$-band  magnitudes 
for sample S4 (solid line) and S3 (dotted line) are similar as well.
}
\end{figure}

\clearpage\setcounter{figure}{3}
\begin{figure}
\epsscale{1}
\plotone{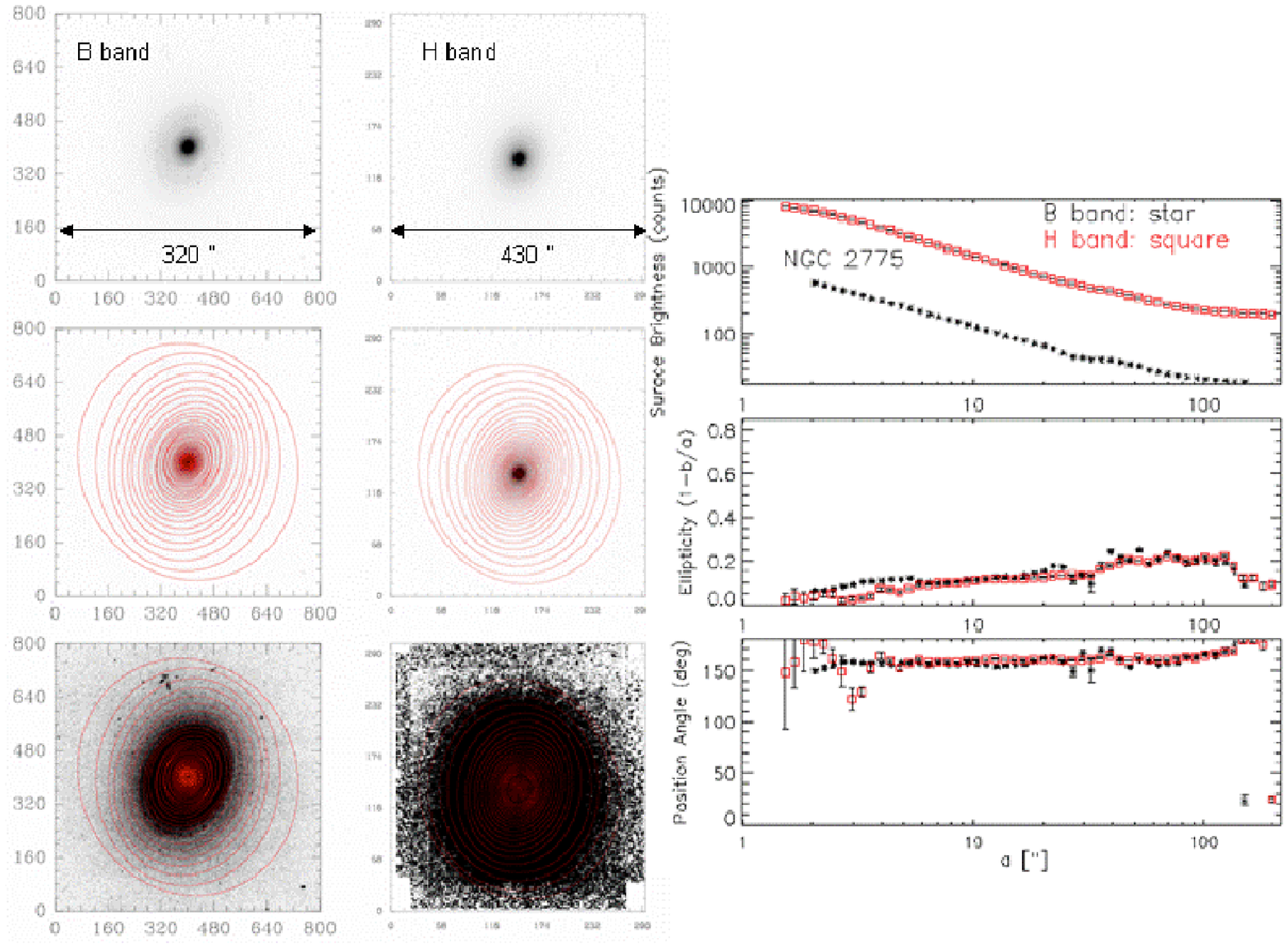}
\caption{\bf 
Example of ellipse fits for the unbarred galaxy NGC 2775:
\rm
Left and middle panels:  They  show the ellipse-fits overlayed 
on the $B$- and $H$-band images of the unbarred 
galaxy NGC~2775. The scales of the $B$ and $H$ images are shown in the
 top image panels for each band. $1\arcsec$ corresponds to 86 pc at the galaxy 
distance of 17 Mpc. 
Within each panel, there are 3 images with different
greyscale  stretches that are chosen to  emphasize the inner  
(middle image) and outer (bottom image) regions of the 
galaxy. Note that ellipses are fitted out to the sky level in the 
image.
Right panel: This shows the radial profiles of (SB, $e$, and PA)
for the $B$ (stars) and $H$ (squares) bands, derived from the 
ellipse fits prior to deprojection.
The profiles  do not show any characteristic bar 
signatures, such as a smooth rise in $e$ to a maximum above 0.25, 
concurrent with a PA plateau. The $e$ remains below 0.25 across the
galaxy. There is no signature of large-scale 
structure, such as spiral arms or a bar. 
} 
\end{figure}

\clearpage\setcounter{figure}{4}
\begin{figure}
\epsscale{1}
\plotone{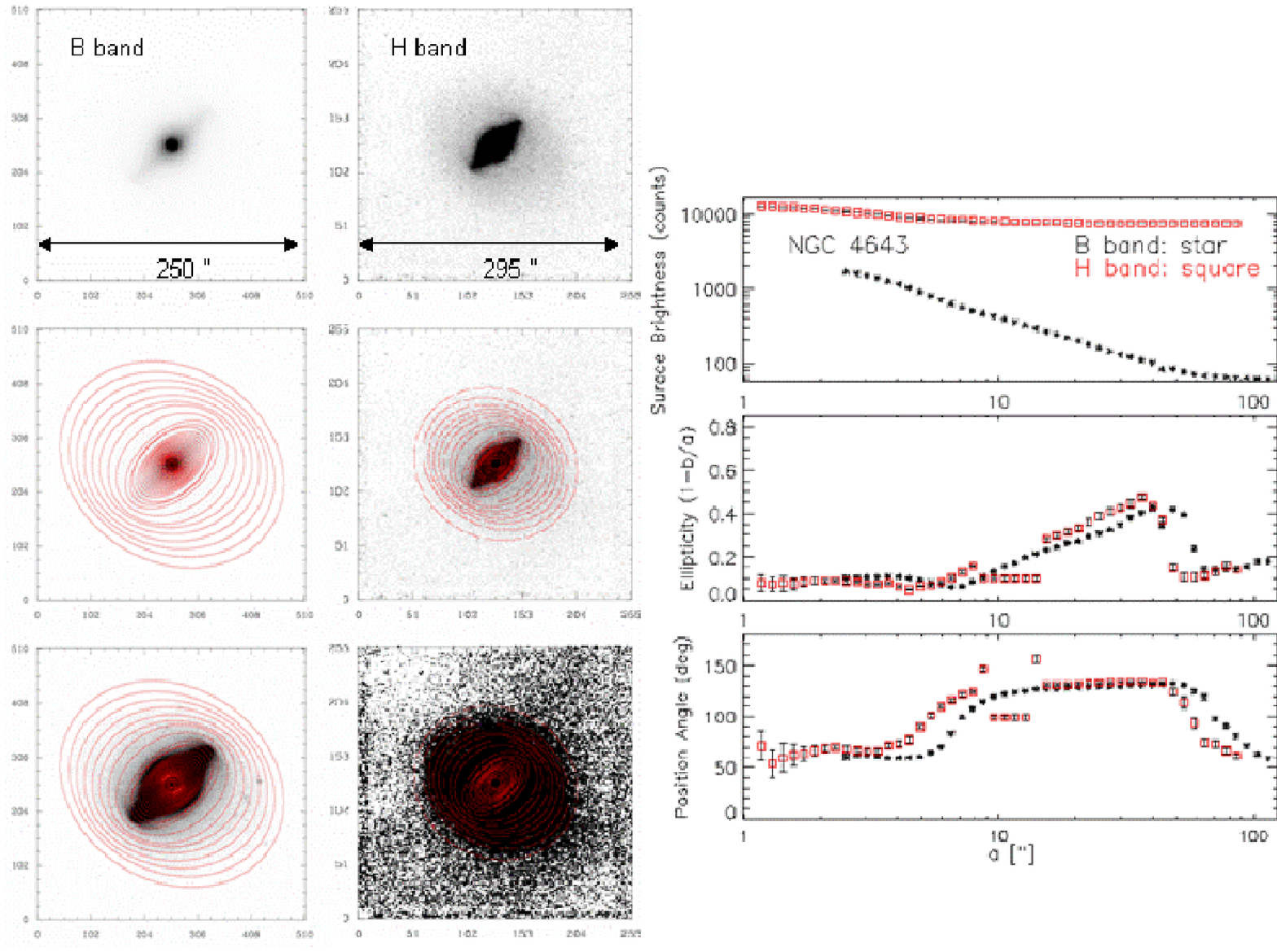}
\caption{\bf 
Example of ellipse fits for the barred galaxy NGC 4643:
\rm
Left and middle panels:  They  show the ellipse-fits overlayed 
on the $B$- and $H$-band images of the barred 
galaxy NGC~4643. The scales of the $B$ and $H$ images are shown in the
 top image panels for each band. $1\arcsec$ corresponds to 130 pc at the galaxy distance of 26 Mpc. 
Within each panel, there are 3 images with different
greyscale  stretches that are chosen to  emphasize the inner  
(middle image) and outer (bottom image) regions of the 
galaxy. Note that ellipses are fitted out to the sky level in the 
image.
Right panel: This shows the radial profiles of (SB, $e$, and PA)
for the $B$ (stars) and $H$ (squares) bands,  derived from the 
ellipse fits and  prior to deprojection. The profiles show a 
clear bar signature. Between 15$\arcsec$ and 40$\arcsec$, the
$e$ rises smoothly to a global maximum of 0.5, while the PA remains
roughly constant. The $e$ then drops to $\sim$ 0.1, and the PA changes
at the
transition from the bar to the disk region.} 
\end{figure}

\clearpage\setcounter{figure}{5}
\begin{figure}
\epsscale{1}
\plotone{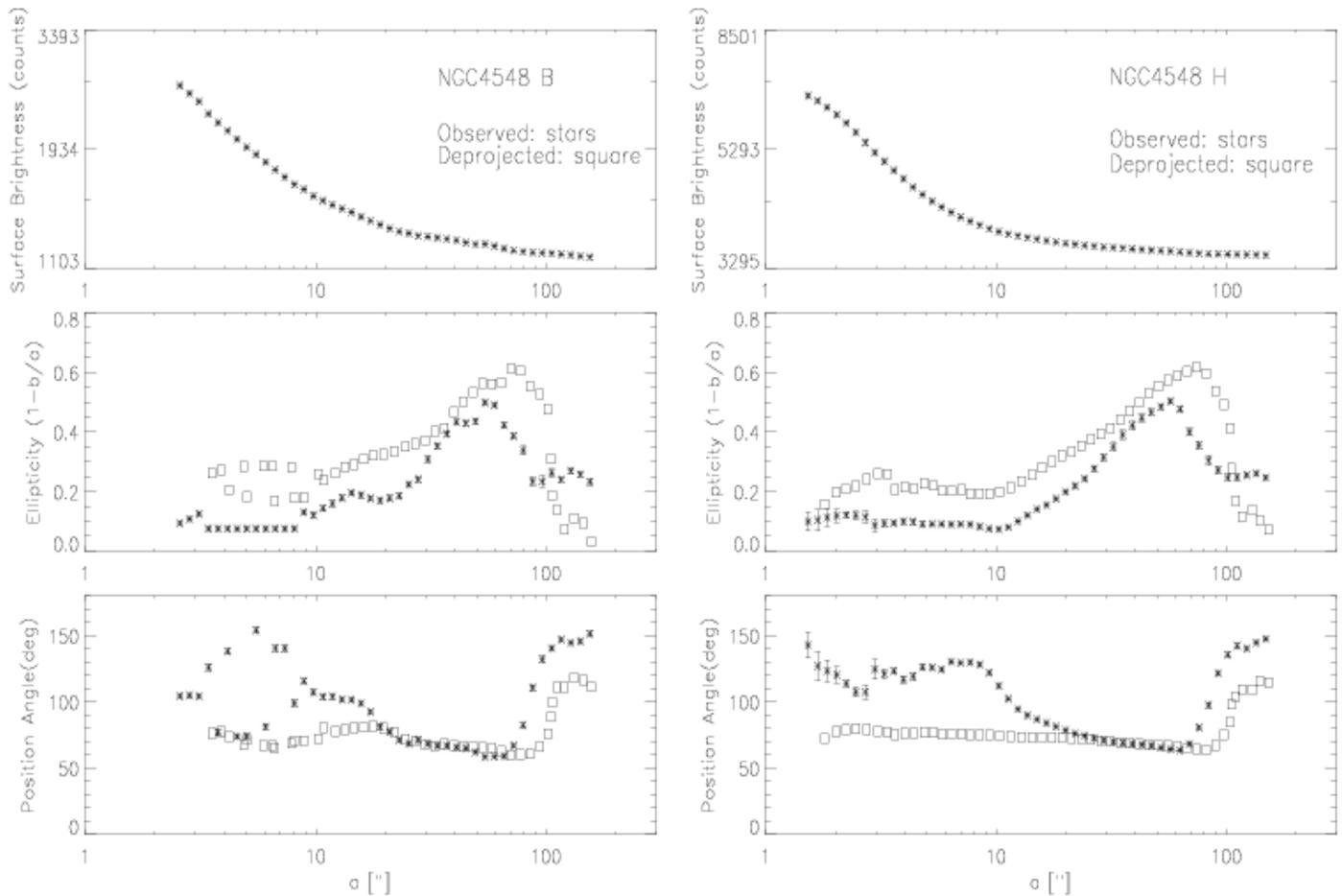}
\caption{\bf 
Example of observed and deprojected radial profiles for NGC 4548: 
\rm
For galaxies in sample S4, we use the inclination $i$ and the PA of the 
outer disk (from $\S$ 3.2) to analytically deproject the observed 
$H$- and $B$-band  radial profiles of ($e$, PA) to face-on. 
The case for NGC~4548 is illustrated here.  
The left panel shows the observed (stars) and deprojected (squares)
radial profiles in the $B$ band. The right panel shows the observed
and deprojected radial profiles in the $H$ band. After deprojection, 
as expected, the  outer disk $e$ is nearly zero in the $B$ band. 
Note also  that the bar size is slightly different and the bar appears 
somewhat stronger in both bands after deprojection.} 
\end{figure}

\clearpage\setcounter{figure}{6}
\begin{figure}
\epsscale{1}
\plotone{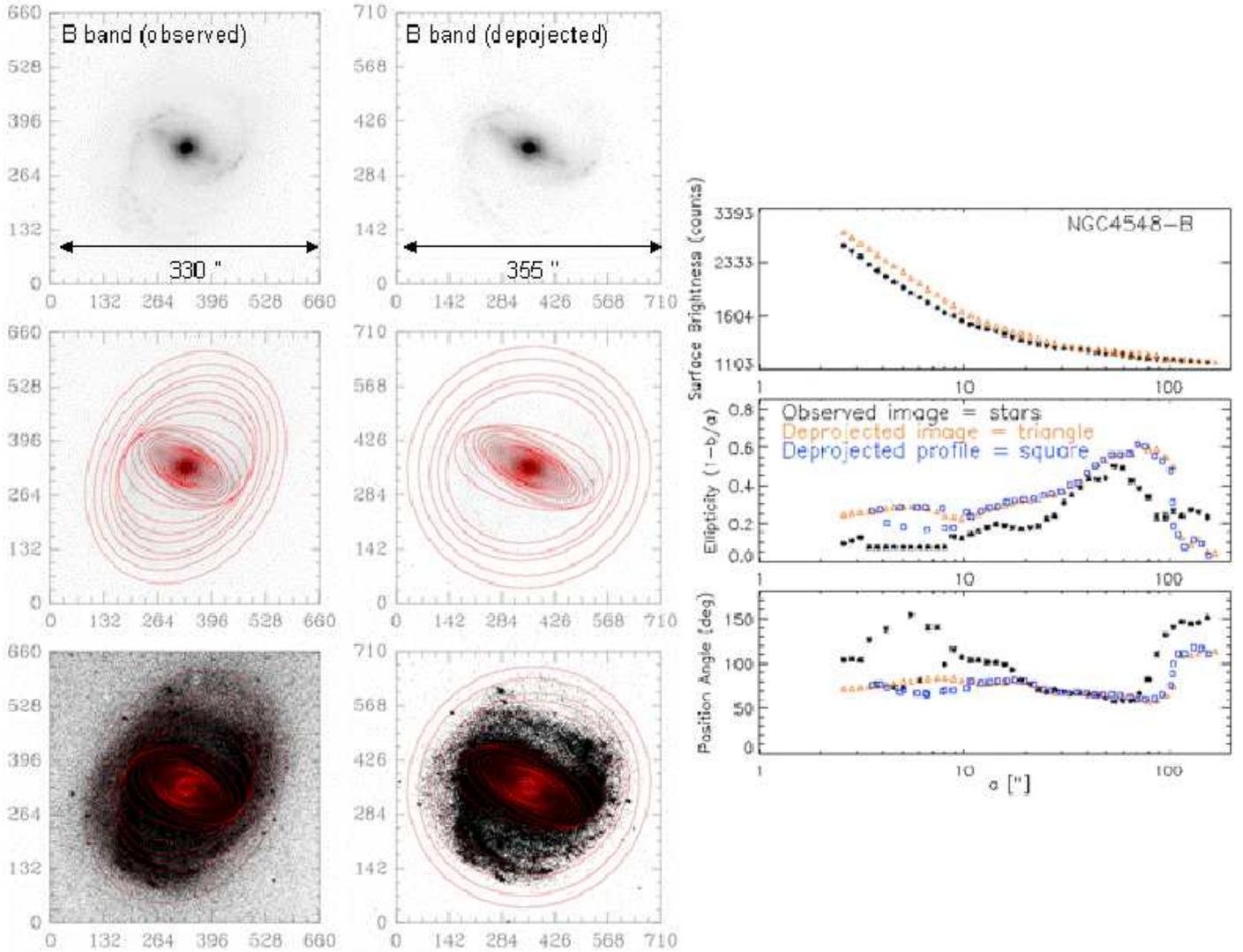}
\caption{\bf 
Comparison of the face-on radial profile generated via 
two different methods:
\rm 
For the $B$-band image of NGC 4548, this figure compares the 
face-on radial profiles of $e$ and PA  generated via two 
different methods. 
In the first method, ellipses are fitted to the observed image 
(left panel) to generate the observed radial profile 
(plotted as stars in the right panel), which is then 
analytically  deprojected to produce the face-on 
profile (plotted  as squares in the right panel).
In the second method, the observed image is deprojected with 
MIRIAD and the resulting deprojected image (middle panel) is fitted
with ellipses to generate the second face-on profile  
(plotted  as triangles in the right panel). Note the good 
agreement between the squares and triangles. } 
\end{figure}

\clearpage
\begin{figure}
\setcounter{figure}{7}
\epsscale{.80}
\plotone{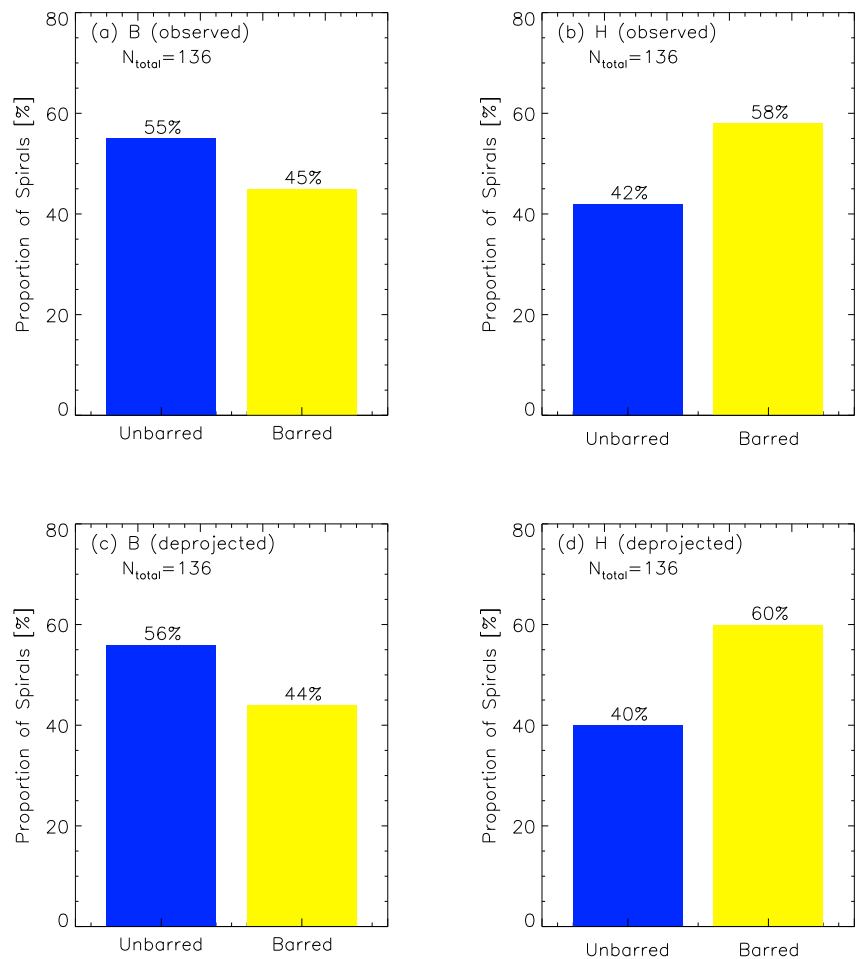} 
\caption{
\bf
The optical and NIR bar fraction at $z\sim$~0 from OSUBSGS: 
\rm
We show the fraction of spirals that are barred in the $B$ and $H$ 
bands, based on ellipse fits of  136 moderately inclined galaxies  
(sample S4), followed by quantitative characterization of 
the resulting radial  profiles of ($e$, SB, PA).
Top row: The observed bar fraction before deprojection 
is  45\% in the $B$ band (left) and 58\% in the $H$ band (right).
Bottom row: The deprojected bar fraction  is 44\% in the $B$ band
(left)  and 60\% in the $H$ band (right).
}
\end{figure}

\clearpage
\begin{figure}
\setcounter{figure}{8}
\epsscale{.90}
\plotone{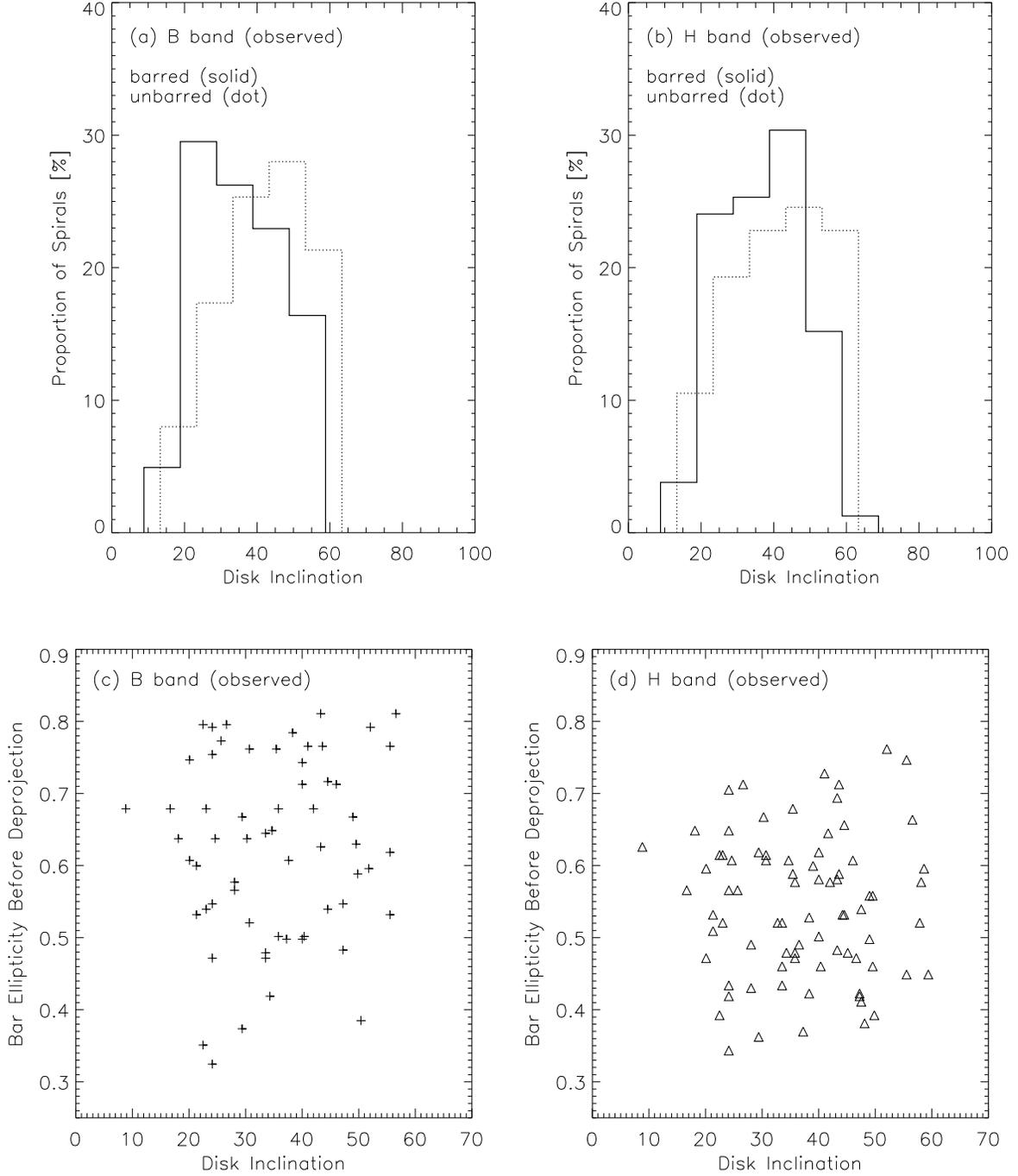}
\caption{ 
\bf 
Verifying that bar properties measured prior to deprojection are not 
biased by galaxy inclination:
\rm
Top tow: The histograms show the distributions of inclination $i$ 
for galaxies  that were classified as `barred' or `unbarred', prior 
to deprojection, in the $B$ band (left) and  $H$ band (right). 
Note that  there is no correlation with $i$. 
Bottom row:  The measured bar ellipticity $e_{\rm bar}$  in 
the  $B$ band (left) and $H$ band (right), prior to deprojection,  
are plotted against  the galaxy inclination $i$. 
Note that there is no correlation between  $e_{\rm bar}$ and $i$.
}
\end{figure}

\clearpage
\begin{figure}
\setcounter{figure}{9}
\epsscale{.80}
\plotone{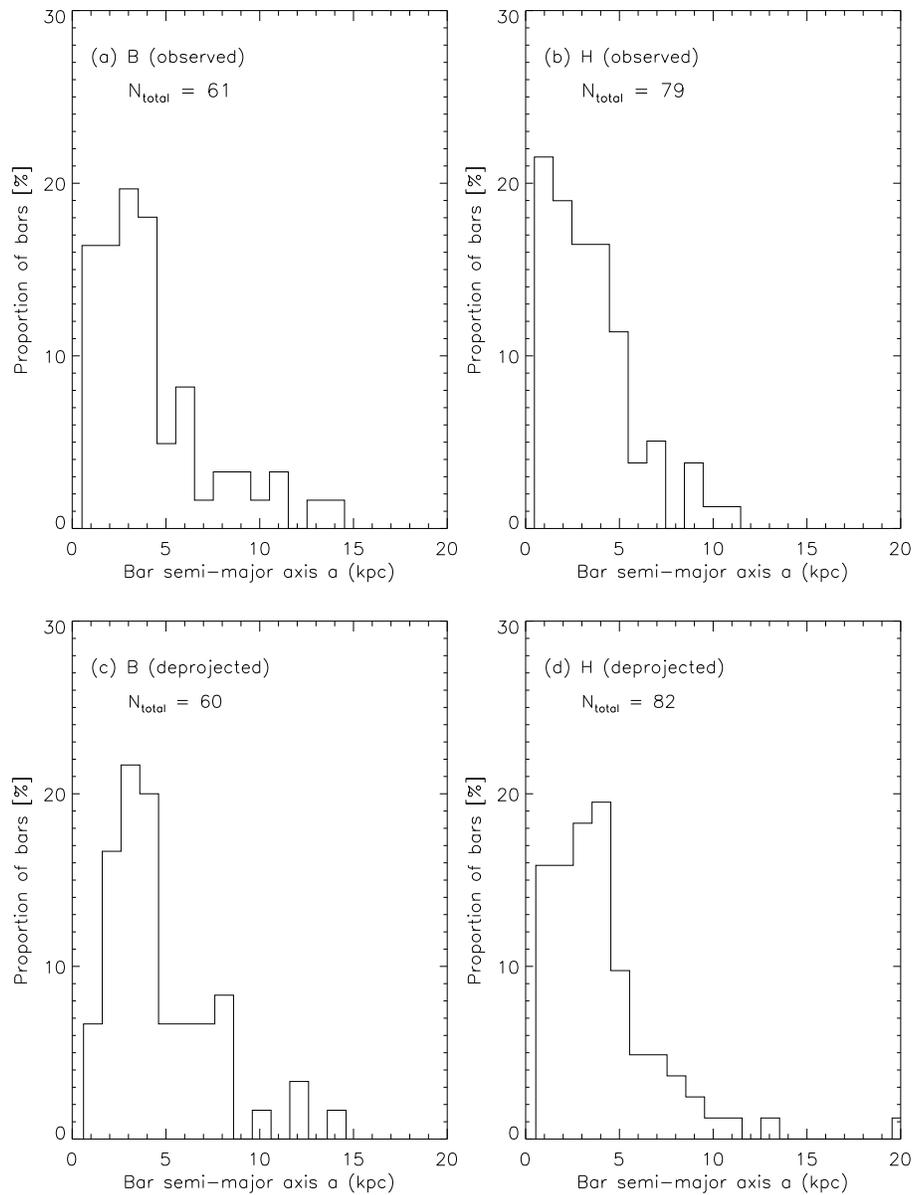}
\caption{
\bf 
Distribution of bar sizes at $z\sim$~0 from OSUBSGS: 
\rm
The distributions of bar semi-major axes ($a_{\rm bar}$) 
before (top row) and after (bottom row) deprojection are
shown, for the $B$ (left) and $H$ (right)  bands. 
Most  (68\% in $B$ and 76\% in $H$ ) bars have $a_{\rm bar} \le$ 5 kpc,
and $\sim$~50\% of them cluster in the range  2 to 5 kpc.
Deprojection makes several bars appear 
somewhat larger, but does not otherwise 
produce a large change in the overall shape of the 
distributions.
}
\end{figure}

\clearpage
\begin{figure}
\setcounter{figure}{10}
\epsscale{.90}
\plotone{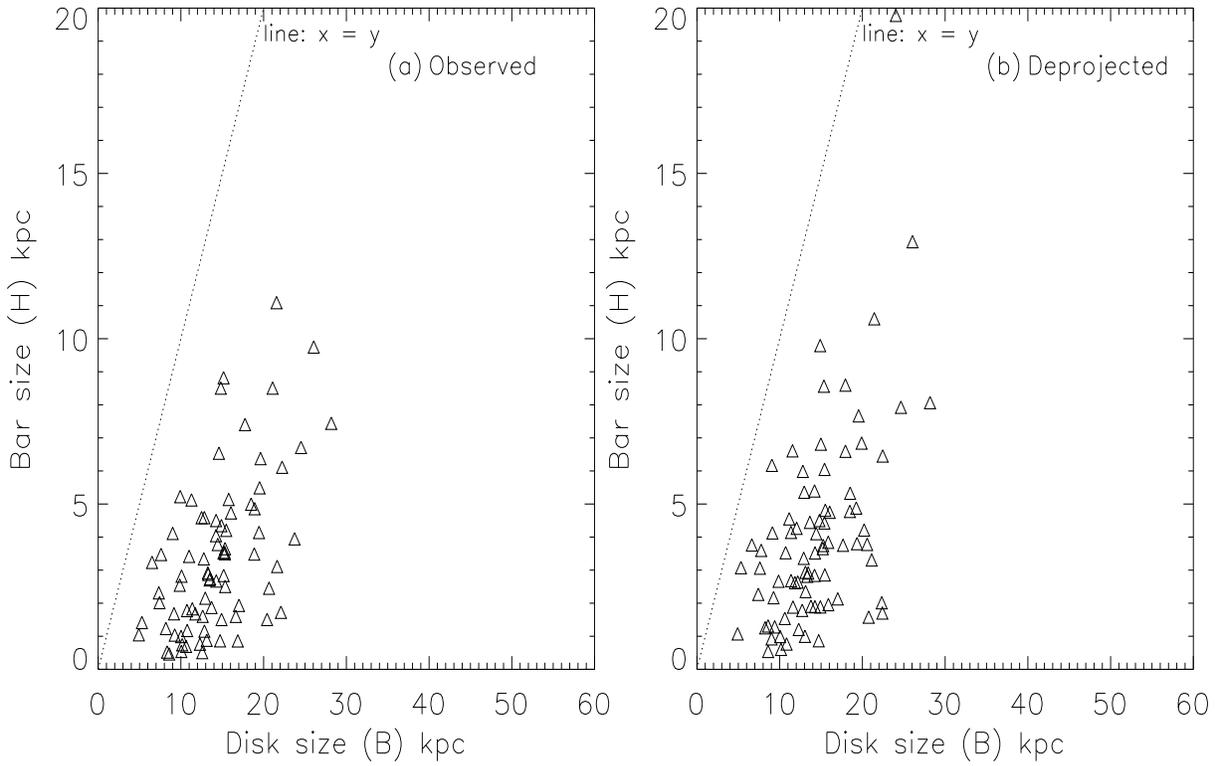}
\caption{
\bf
Relationship between $H$-band bar size and disk size 
at $z\sim$~0 from OSUBSGS: 
\rm 
The bar semi-major axis  in the $H$ band is plotted versus 
the disk size  before (left panel) and after (right panel) deprojection.
The disk size is  measured in the $B$-band image which is 
deeper than the $H$ band and traces the disk further out.
The deprojected bar and disk sizes are correlated  
with an average slope of $\sim$~0.9. However, there is 
a large scatter of several kpc in bar size at a given 
disk size. For comparison, the dotted line has slope of 1. 
}
\end{figure}

\clearpage
\begin{figure}
\setcounter{figure}{11}
\epsscale{.90}
\plotone{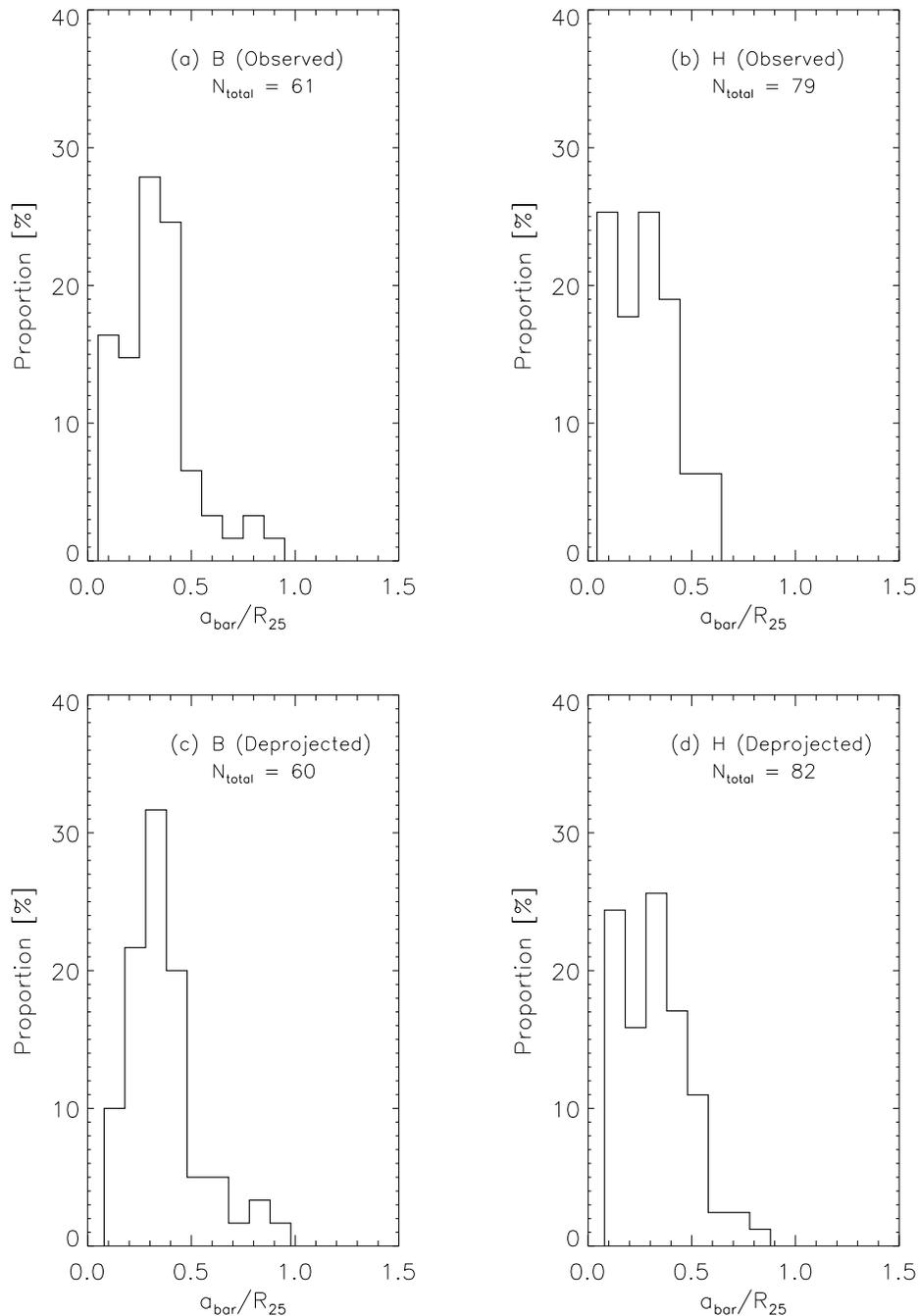}
\caption{
\bf 
Relationship between bar size and  $R_{\rm 25}$ at $z\sim$~0 from OSUBSGS:
\rm 
The ratio  of the bar semi-major axis  ($a_{\rm bar}$)
to the isophotal  radius ($R_{\rm 25}$)  where the $B$-band 
surface brightness is 25 mag arcsec$^{-2}$ is shown  
before (top row) and after (bottom row) deprojection.
In the left panels, the bar size ($a_{\rm bar}$) is 
determined from the $B$-band image and in the right panels
from the $H$-band image.
We find that 
the ratio ($a_{\rm bar}$/$R_{\rm 25}$) is always below 1.0, and 
lies primarily in the range 0.2 to 0.4 in both  $H$ and $B$ bands.
}
\end{figure}

\clearpage
\begin{figure}
\setcounter{figure}{12}
\epsscale{.90}
\plotone{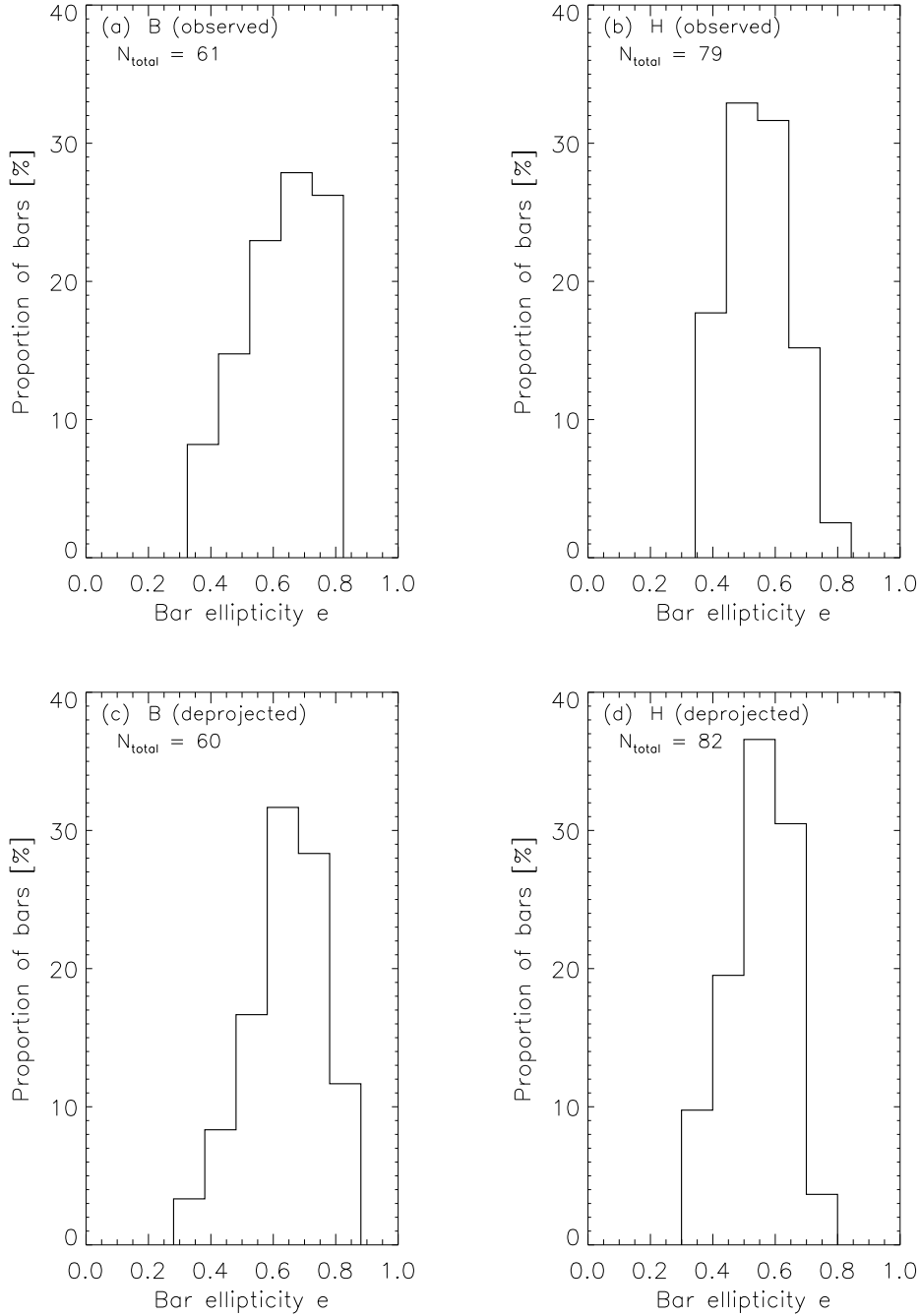}
\caption{
\bf 
Distributions of bar strengths as characterized by  ellipse-fitting   
at $z\sim$~0 from OSUBSGS: 
\rm
The distributions of bar strengths  
(as characterized by  $e_{\rm bar}$ from ellipse-fitting) before 
(top row) and after (bottom row)  deprojection,  in 
the $B$ (left) and $H$  (right) bands  are shown.
It is striking that only a tiny fraction 
(7\% in $B$; 10\% in $H$) of bars  are very weak 
with $e_{\rm bar}$  between 0.25--0.40, while 
the majority of bars  (70\% in $B$; 71\% in $H$)  
seem to have moderate to high ellipticities, with 
$e_{\rm bar}$  between 0.50 to 0.75. 
Furthermore,  we find no evidence for  bimodality in the 
distribution of bar strength as characterized by  $e_{\rm bar}$ 
in the $B$ or $H$ bands. 
} 
\end{figure}


\clearpage
\begin{figure}
\setcounter{figure}{13}
\epsscale{.80}
\plotone{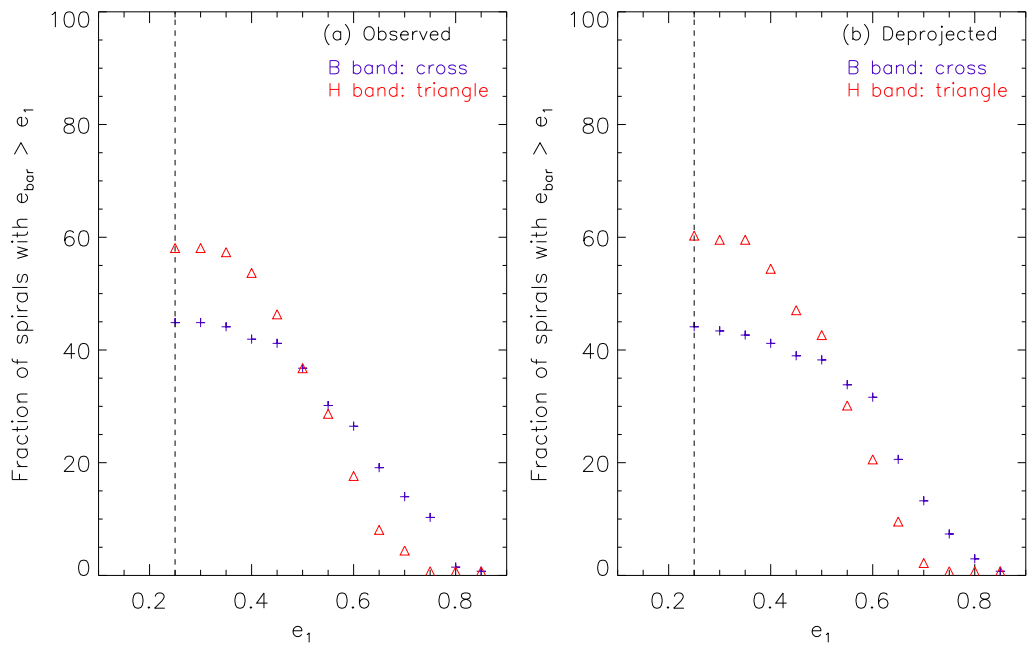} 
\caption{
\bf 
Bar fraction as a function of bar strength as characterized by  $e_{\rm bar}$: 
\rm
A generalized plot of the fraction  
of disks with `strong' and `weak' bars is shown 
before  (left panel) and after (right panel) deprojection.
The bar strength here is characterized by  $e_{\rm bar}$ from ellipse-fitting.
The y-axis   shows  the fraction of spiral galaxies that  host bars 
whose strength $e_{\rm bar}$ exceeds a value $e_{\rm 1}$  in the 
$B$ (cross) and $H$ (triangle) bands. Along the the x-axis, $e_{\rm 1}$ 
is varied. 
As $e_{\rm 1}$  rises from 0.35 to 0.45, 0.55, and 0.75, 
the deprojected bar fraction  in the $B$ band falls  from 
43\% to  39\%, 34\%, and  7\%, respectively, 
while the  bar fraction  in the $H$ band falls from 
59\% to 47\%, 30\%, and  1\%.
The flattening of the curve around  $e_{\rm 1} \sim$~0.45  
reflects  the paucity of very weak (low ellipticity) bars 
with 0.25~$ \ge  e_{\rm bar}  
\le$~0.40, while the steep fall in the curve  for  $e_{\rm 1}$
in the range 0.50--0.75 shows the preponderance of `strong' 
(high ellipticity) bars. 
}
\end{figure}

\clearpage
\begin{figure}
\setcounter{figure}{14}
\epsscale{.8}
\plotone{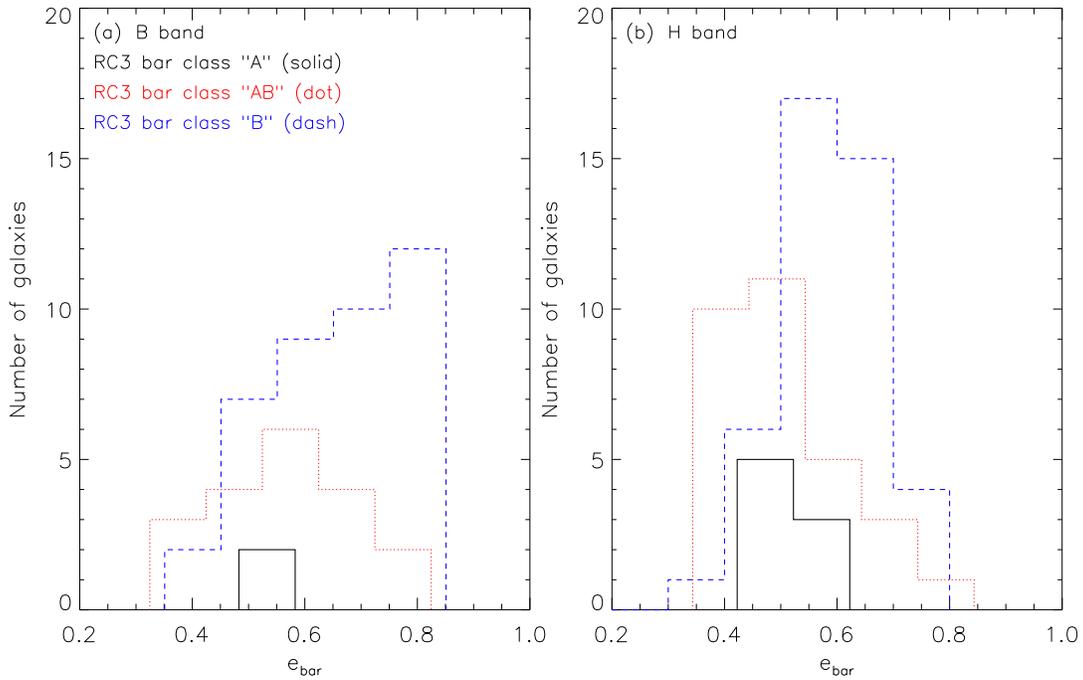} 
\vskip -0.5 in 
\caption{
\bf
A comparison of RC3 visual bar classes with $e_{\rm bar}$ from ellipse fits:
\rm 
This figure shows the RC3 visual bar classes   
for all those galaxies in sample S4  that we classified as barred 
based on ellipse fits ($\S$~3.3 and $\S$~3.4).
The x-axis shows the  bar strength as characterized 
by  $e_{\rm bar}$ from ellipse-fitting in  the $B$ (left panel) and $H$ 
(right panel) bands, prior to deprojection.
The three  RC3  visual  bar classes are based on  visual inspection 
of optical images (de Vaucouleurs \etal 1991)  and classes 
 `A' (solid line), `AB' (dotted line), and `B' (dashed line) denote 
`unbarred', `weakly barred', and  `strongly barred' disks, respectively.
In the $B$ band, we find that   5\%, 41\%, and 85\%, respectively, 
of  the sample galaxies with RC3 visual classes of `A', `AB', and `B', 
host bars.  
In the $H$ band,  the corresponding numbers are 19\%, 65\%, 
and 87\%, respectively. Thus, many  galaxies that are classified
as unbarred in RC3 turn out to be barred and vice-versa.  
The mean bar ellipticity  $e_{\rm  bar}$  is higher 
for RC3 visual class ``B'' than for class ``AB'', but 
the two classes have significant overlap in the range 
$e_{\rm  bar} \sim$~0.5--0.7.
}
\end{figure}

\clearpage
\begin{figure}
\setcounter{figure}{15}
\epsscale{.70}
\plotone{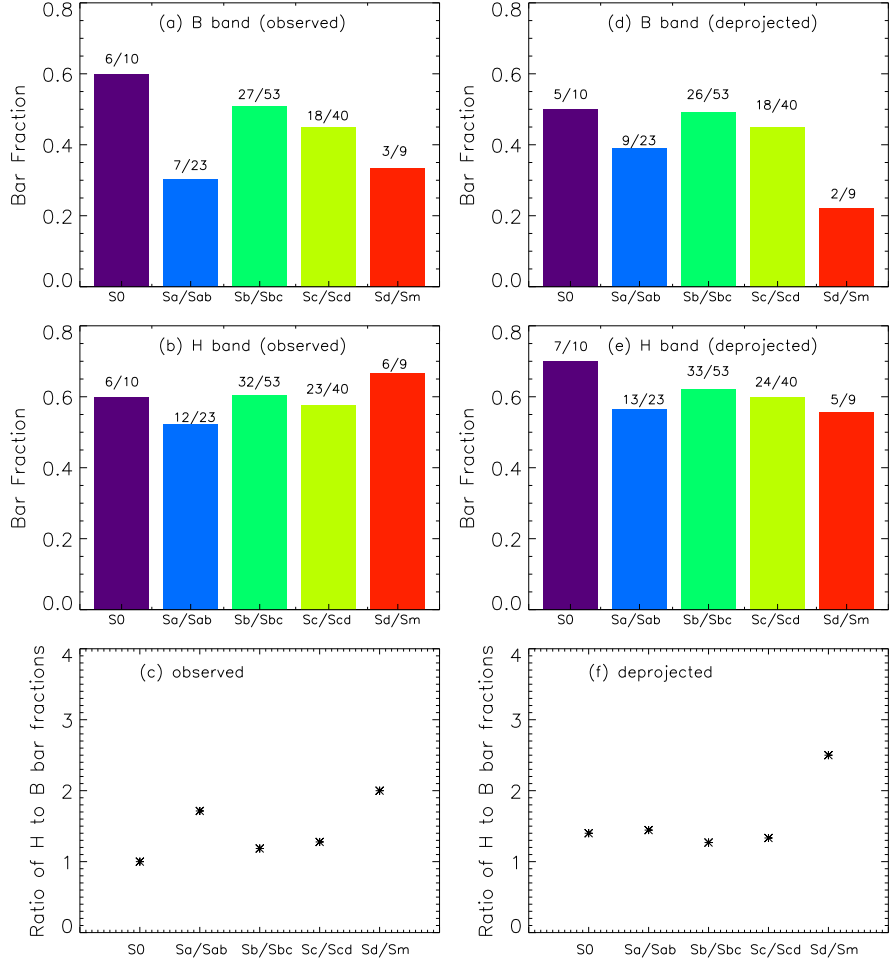}  
\caption{
\bf 
Bar fraction as a function of RC3 Hubble types at $z\sim$~0 from OSUBSGS: 
\rm
This figure  shows what proportion of spirals with different 
Hubble types host bars.
The bar fractions in  the $B$ band (top row) and  
$H$ band  (middle row) are shown as a function of RC3 Hubble 
types, before (left) and after (right) deprojection. 
The bar fraction  above each bin is explicitly given as 
the ratio (number of barred disks with a given Hubble type/total  
number of disks of a given Hubble type).
The number of galaxies are small for S0 and Sd/Sm types and robust 
number statistics only apply to  RC3 Hubble types Sa to Scd: 
we find that the $H$-band bar fraction   remains at  $\sim$ 60\%  
across RC2 Hubble types Sa to Scd. 
The bottom row shows the ratio of the $H$-band bar fraction 
to the $B$-band bar fraction  before (left) and  after (right) 
deprojection. 
In  the $B$ band, we find that the bar fraction is lower with 
respect to the $H$ band by  ~$\sim$~1.2--1.5  for S0s to Scs, 
and by $\sim$~2.5  for Sds/Sms. This  is likely due to extinction,
especially in the dusty, gas-rich  late type (Scd--Sm) galaxies.
}
\end{figure}

\clearpage
\begin{figure}
\setcounter{figure}{16}
\epsscale{.70}
\plotone{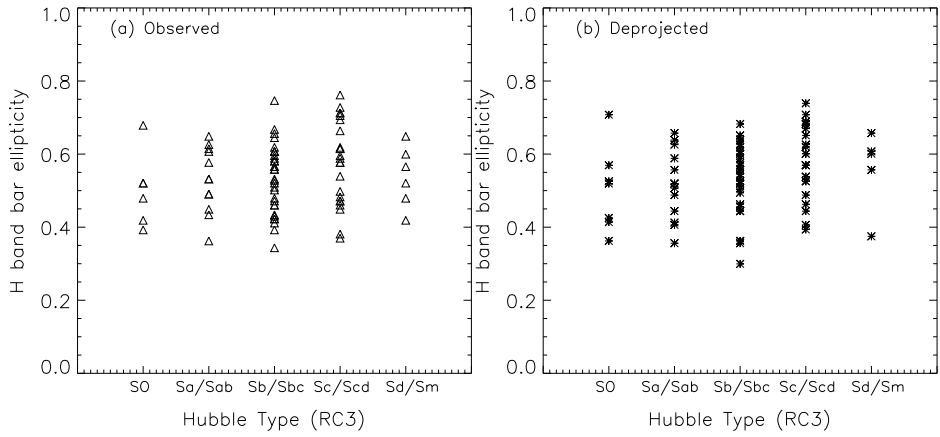}   
\vskip -0.5 in
\caption{
\bf 
Bar strength  as characterized by  $e_{\rm bar}$ 
as a function of RC3 Hubble types  at $z\sim$~0 from OSUBSGS:
\rm
 The bar strength as characterized by the  bar ellipticity 
$e_{\rm bar}$ in the $H$ band
 is plotted as a function of  Hubble types  before (left panel)
 and after (right panel)  deprojection. The Hubble types are from
 RC3 and are binned as in Figure  16.
 Before deprojection, the number of galaxies
 in each Hubble type bin is: SO~=~6, Sa/Sab~=~12, Sb/Sbc~=~32,
 Sc/Scd~=~23, Sd/Sm~=6.
 After deprojection, the corresponding numbers are
 SO~=~7, Sa/Sab~=~13, Sb/Sbc~=~33, Sc/Scd~=~24, Sd/Sm~=~5.
 The number of galaxies are small for S0s and Sd/Sm types and robust 
 number statistics only apply to  RC3 Hubble types Sa to Scd.
 The  bar ellipticity  $e_{\rm bar}$ lies in
 the range 0.35--0.80, and shows no systematic variation across
 Hubble types  Sa to Scd, either before or after  deprojection.
} 
\end{figure}


%
%
\clearpage
\setcounter{table}{0}
\begin{deluxetable}{lcccccccc}
\tabletypesize{\scriptsize} 
\tablewidth{0pt}
\tablecaption{Global Properties of sample S3 (169 galaxies) with ellipse 
fits in $B$ and $H$}
\tablehead{
\colhead {Galaxy Name} & 
\colhead {Hubble Type} &
\colhead {Bar Type} &
\colhead { $D$ } &
\colhead { $D_{\rm\tiny 25}$  } &
\colhead { $B_{\rm \tiny T}$  } &
\colhead { $M_{\rm \tiny V}$  } &
\colhead { $L_{\rm \tiny IR}$  }&
\colhead { $L_{\rm \tiny B}$  } \\
\colhead{}   & 
\colhead{ (RC3) }   & 
\colhead{ (RC3) }   & 
\colhead{ (Mpc)  }   & 
\colhead{ (')  }   & 
\colhead{ (mag) }   &
\colhead{ (mag) }   &
\colhead{ (log(\mbox{L$_{\odot}$})) } &
\colhead{ (log(\mbox{L$_{\odot}$})) }  \\
\colhead {(1)} & \colhead {(2)} & \colhead {(3)} & \colhead {(4)} &
\colhead {(5)} & \colhead {(6)} & \colhead {(7)} & 
\colhead {(8)}  & \colhead {(9)}\\
}
\startdata
 & & & & & & & & \\
\multicolumn {9}{c} {\small Moderately inclined galaxies (N=136)} \\
\hline \\
IC 0239    &    SAB(rs)cd    & AB &14.2&5.4&11.8&-19.66&  -  & 9.81\\ 
  IC 4444    &    SAB(rs)bc    & AB &26.9&1.4&12&-20.79&10.53& 10.33\\ 
  IC 5325    &    SAB(rs)bc    & AB &18.1&2.7&11.83&-20.02&  -  & 9.83\\ 
  NGC 0157    &    SAB(rs)bc    & AB &20.9&3&11&-21.19&10.52& 10.53\\ 
  NGC 0210    &    SAB(s)b    & AB &20.3&4.9&11.6&-20.65&  -  & 10.22\\ 
  NGC 0278    &    SAB(rs)b    & AB &11.8&2.7&11.47&-19.53&10.03& 10.04\\ 
  NGC 0289    &    SAB(rs)bc    & AB &19.4&8.3&11.72&-20.45&10.03& 10.17\\ 
  NGC 0428    &    SAB(s)m    & AB &14.9&4.6&11.91&-19.4&  -  & 9.85\\ 
  NGC 0488    &    SA(r)b    & A &29.3&5.4&11.15&-22.05&  -  & 10.74\\ 
  NGC 0685    &    SAB(r)c    & AB &15.2&3.9&11.95&-19.42&  -  & 9.8\\ 
  NGC 0864    &    SAB(rs)c    & AB &20&4.4&11.4&-20.66&  -  & 10.27\\ 
  NGC 1042    &    SAB(rs)cd    & AB &16.7&4.4&11.56&-20.09&  -  & 10.16\\ 
  NGC 1058    &    SA(rs)c    & A &9.1&3.6&11.82&-18.6&  -  & 9.34\\ 
  NGC 1073    &    SB(rs)c    & B &15.2&5&11.47&-19.94&  -  & 9.97\\ 
  NGC 1084    &    SA(s)c    & A &17.1&3.4&11.31&-20.43&10.54& 10.3\\ 
  NGC 1087    &    SAB(rs)c    & AB &19&3.7&11.46&-20.45&10.26& 10.28\\ 
  NGC 1187    &    SB(r)c    & B &16.3&5.4&11.34&-20.28&10.18& 10.1\\ 
  NGC 1241    &    SB(rs)b    & B &26.6&3.6&11.99&-20.98&  -  & 10.12\\ 
  NGC 1300    &    SB(rs)bc    & B &18.8&6.8&11.11&-20.94&  -  & 10.36\\ 
  NGC 1302    &    (R)SB(r)0    & B &20&3.8&11.6&-20.8&  -  & 10.24\\ 
  NGC 1309    &    SA(s)bc    & A &26&2.8&11.97&-20.54&10.24& 10.26\\ 
  NGC 1317    &    SAB(r)a    & AB &16.9&3.1&11.91&-20.12&  -  & 9.87\\ 
  NGC 1350    &    (R')SB(r)ab    & B &16.9&5&11.16&-20.85&  -  & 10.18\\ 
  NGC 1371    &    SAB(rs)a    & AB &17.1&6.8&11.57&-20.49&  -  & 10.08\\ 
  NGC 1385    &    SB(s)cd    & B &17.5&4.6&11.45&-20.28&10.18& 10.1\\ 
  NGC 1493    &    SB(r)cd    & B &11.3&3.9&11.78&-19&  -  & 9.58\\ 
  NGC 1559    &    SB(s)cd    & B &14.3&3.3&11&-20.13&10.21& 10.24\\ 
  NGC 1617    &    SB(s)a    & B &13.4&4&11.38&-20.2&  -  & 10.08\\ 
  NGC 1637    &    SAB(rs)c    & AB &8.9&5.1&11.47&-18.92&9.46& 9.52\\ 
  NGC 1703    &    SB(r)b    & B &17.4&3.5&11.9&-19.86&  -  &  - \\ 
  NGC 1792    &    SA(rs)bc    & A &13.6&6.1&10.87&-20.48&10.33& 10.24\\ 
  NGC 1832    &    SB(r)bc    & B &23.5&2.6&11.96&-20.53&10.28& 10.26\\ 
  NGC 2139    &    SAB(rs)cd    & AB &22.4&2.9&11.99&-20.12&10.16& 10.16\\ 
  NGC 2196    &    (R')SA(s)a    & A &28.8&2.9&11.82&-21.29&  -  & 10.46\\ 
  NGC 2566    &    (R')SB(rs)ab pec    & B &21.1&4.3&11.83&-20.6&10.6&  -\\  
  NGC 2775    &    SA(r)ab    & A &17&4.6&11.03&-21.02&  -  & 10.24\\ 
  NGC 2964    &    SAB(r)bc    & AB &21.9&3&11.99&-20.39&10.36& 10.15\\ 
  NGC 3166    &    SAB(rs)0    & AB &22&3.2&11.32&-21.32&9.94& 10.28\\ 
  NGC 3169    &    SA(s)a pec    & A &19.7&5&11.08&-21.24&10.18& 10.37\\ 
  NGC 3223    &    SA(s)b    & A &38.1&3.6&11.79&-21.93&  -  & 10.88\\ 
  NGC 3227    &    SAB(s)a pec    & AB &20.6&5.9&11.1&-21.29&10.13& 10.26\\ 
  NGC 3261    &    SB(rs)b    & B &33.4&3.9&12&  -  &  -  & 10.64\\ 
  NGC 3275    &    SB(r)ab    & B &42.4&2.8&11.8&  -  &  -  & 10.61\\ 
  NGC 3423    &    SA(s)cd    & A &10.9&4&11.59&-19.05&  -  & 9.66\\ 
  NGC 3504    &    (R)SAB(s)ab    & AB &26.5&2.6&11.82&-20.99&10.72& 10.34\\ 
  NGC 3507    &    SB(s)b    & B &19.8&3.2&11.73&  -  &  -  & 10.23\\ 
  NGC 3513    &    SB(rs)c    & B &17&3.2&11.93&-19.65&  -  & 9.98\\ 
  NGC 3583    &    SB(s)b    & B &34&2.6&11.9&  -  &10.54& 10.61\\ 
  NGC 3593    &    SA(s)0    & A &5.5&4.8&11.86&-17.78&9.22& 9\\ 
  NGC 3596    &    SAB(rs)c    & AB &23&4.1&11.95&  -  &  -  & 10.32\\ 
  NGC 3646    &    Ring    &  -  &55.8&3.9&11.78&-22.6&  -  &  - \\ 
  NGC 3681    &    SAB(r)bc    & AB &24.2&2.9&11.9&-20.73&  -  & 10.06\\ 
  NGC 3684    &    SA(rs)bc    & A &23.4&2.9&12&-20.47&  -  & 10.07\\ 
  NGC 3686    &    SB(s)bc    & B &23.5&2.9&11.89&-20.54&  -  & 10.17\\ 
  NGC 3726    &    SAB(r)c    & AB &17&5.5&10.91&-20.73&9.78& 10.33\\ 
  NGC 3810    &    SA(rs)c    & A &16.9&3.8&11.35&-20.37&10.12& 10.24\\ 
  NGC 3885    &    SA(s)0    & A &27.8&2.9&11.89&-21.28&10.27& 10.29\\ 
  NGC 3887    &    SB(r)bc    & B &19.3&3.4&11.41&  -  &9.8& 10.16\\ 
  NGC 3893    &    SAB(rs)c    & AB &17&4.3&11.16&  -  &10.2& 10.3\\ 
  NGC 3938    &    SA(s)c    & A &17&4.9&10.9&-20.77&9.93& 10.3\\ 
  NGC 3949    &    SA(s)bc    & AB &17&2.8&11.54&-20.06&9.87& 10.16\\ 
  NGC 4027    &    SB(s)dm    & B &25.6&3.3&11.66&-20.92&10.36& 10.41\\ 
  NGC 4030    &    SA(s)bc    & A &25.9&4&11.42&  -  &10.64& 10.3\\ 
  NGC 4051    &    SAB(rs)bc    & AB &17&5.4&10.83&-20.97&9.9& 10.29\\ 
  NGC 4123    &    SB(r)c    & B &16.5&4.6&11.98&-19.71&9.76& 10.29\\ 
  NGC 4136    &    SAB(r)c    & AB &9.7&4&11.69&  -  &  -  & 9.48\\ 
  NGC 4145    &    SAB(rs)d    & AB &20.7&5.9&11.78&-20.31&  -  & 10.28\\ 
  NGC 4151    &    (R')SAB(rs)ab    & AB &20.3&6.3&11.5&-20.77&10.2& 10.38\\ 
  NGC 4212    &    SAc    & A &16.8&2.3&11.83&-19.97&9.82& 10.02\\ 
  NGC 4242    &    SAB(s)dm    & AB &7.5&5.2&11.37&-18.55&  -  & 9.36\\ 
  NGC 4254    &    SA(s)c    & A &16.8&5&10.44&  -  &10.54& 10.53\\ 
  NGC 4303    &    SAB(rs)bc    & AB &15.2&5.9&10.18&-21.26&10.51& 10.48\\ 
  NGC 4314    &    SB(rs)a    & B &9.7&4.2&11.43&-19.35&  -  & 9.65\\ 
  NGC 4394    &    (R)SB(r)b    & B &16.8&3.4&11.73&-20.25&  -  & 9.99\\ 
  NGC 4414    &    SA(rs)c    & A &9.7&4.5&10.96&-19.81&10.56& 9.84\\ 
  NGC 4450    &    SA(s)ab    & A &16.8&5&10.9&-21.05&  -  & 10.34\\ 
  NGC 4457    &    (R)SAB(s)0    & AB &17.4&3.2&11.76&-20.29&  -  & 10.01\\ 
  NGC 4487    &    SAB(rs)cd    & AB &19.9&3.9&11.63&  -  &  -  & 10.27\\ 
  NGC 4496    &    SB(rs)m    & B &13.1&3.7&11.94&-19.17&  -  & 9.8\\ 
  NGC 4504    &    SA(s)bc    & A &19.5&3.3&11.89&  -  &  -  & 10.17\\ 
  NGC 4548    &    SB(rs)b    & B &16.8&5&10.96&-20.98&  -  & 10.3\\ 
  NGC 4571    &    SA(r)d    & A &16.8&3.6&11.82&-19.82&  -  & 9.94\\ 
  NGC 4579    &    SAB(rs)b    & AB &16.8&5.4&10.48&-21.47&9.87& 10.46\\ 
  NGC 4580    &    SAB(rs)a pec    & AB &25.6&2.5&11.83&  -  &  -  & 9.97\\ 
  NGC 4593    &    (R)SB(rs)b    & B &39.5&3.3&11.67&  -  &  -  & 10.82\\ 
  NGC 4618    &    SB(rs)m    & B &7.3&3.1&11.22&-18.54&  -  & 9.44\\ 
  NGC 4643    &    SB(rs)0    & B &25.7&2.9&11.72&-21.29&  -  & 10.39\\ 
  NGC 4647    &    SAB(rs)c    & AB &16.8&2.8&11.94&-19.84&9.81& 9.93\\ 
  NGC 4651    &    SA(rs)c    & A &16.8&3.6&11.39&-20.31&9.72& 10.22\\ 
  NGC 4665    &    SB(s)0    & B &17.9&4.2&10.5&  -  &  -  & 9.94\\ 
  NGC 4689    &    SA(rs)bc    & A &16.8&3.7&11.6&-20.18&  -  & 10.04\\ 
  NGC 4691    &    (R)SB(s)0 pec    & B &22.5&3.5&11.66&-20.68&10.32& 10.24\\ 
  NGC 4698    &    SA(s)ab    & A &16.8&3.3&11.46&-20.58&  -  & 10.22\\ 
  NGC 4699    &    SAB(rs)b    & AB &25.7&3.1&10.41&-22.53&10.12& 10.89\\ 
  NGC 4775    &    SA(s)d    & A &26.6&2.3&11.67&  -  &  -  & 10.32\\ 
  NGC 4900    &    SB(rs)c    & B &17.3&2.5&11.9&-19.82&9.73& 9.83\\ 
  NGC 4902    &    SB(r)b    & B &39.2&2.6&11.61&-22.05&  -  & 10.65\\ 
  NGC 4930    &    SB(rs)b    & B &35&5.4&12&-21.62&  -  &  - \\ 
  NGC 4939    &    SA(s)bc    & A &44.3&5.6&11.9&-21.97&  -  & 11.16\\ 
  NGC 4941    &    (R)SAB(r)ab    & AB &6.4&4.2&11.9&-17.97&  -  & 9.12\\ 
  NGC 4995    &    SAB(rs)b    & AB &28&2.3&12&-21.11&  -  & 10.4\\ 
  NGC 5005    &    SAB(rs)bc    & AB &21.3&5.6&10.61&-21.83&10.46& 10.7\\ 
  NGC 5054    &    SA(s)bc    & A &27.3&4.6&11.67&-21.27&10.46& 10.66\\ 
  NGC 5085    &    SA(s)c    & A &28.9&3.9&11.96&  -  &  -  & 10.48\\ 
  NGC 5101    &    (R)SB(rs)0    & B &27.4&5.4&11.63&-21.54&  -  & 10.57\\ 
  NGC 5121    &    (R')SA(s)a    & A &22.1&2.2&11.51&-21.16&  -  & 10.08\\ 
  NGC 5247    &    SA(s)bc    & A &22.2&4.6&10.5&-21.77&10.32& 10.57\\ 
  NGC 5334    &    SB(rs)c    & B &24.7&4.2&11.99&  -  &  -  & 10.06\\ 
  NGC 5371    &    SAB(rs)bc    & AB &37.8&4.2&11.32&-22.27&10.67& 10.82\\ 
  NGC 5427    &    SA(s)c pec    & A &38.1&2.3&11.93&-21.54&10.8& 10.57\\ 
  NGC 5483    &    SA(s)c    & A &24.7&4.6&11.93&  -  &10.05& 10.3\\ 
  NGC 5676    &    SA(rs)bc    & A &34.5&4&11.87&-21.5&10.63& 10.77\\ 
  NGC 5701    &    (R)SB(rs)0    & B &26.1&4.2&11.76&-21.2&  -  & 10.33\\ 
  NGC 5713    &    SAB(rs)bc pec    & AB &30.4&3.1&11.84&-21.21&10.72& 10.43\\ 
  NGC 5850    &    SB(r)b    & B &28.5&4.6&11.54&-21.52&  -  & 10.47\\ 
  NGC 5921    &    SB(r)bc    & B &25.2&4.9&11.49&-21.18&  -  & 10.46\\ 
  NGC 5962    &    SA(r)c    & A &31.8&2.8&11.98&-21.17&10.55& 10.46\\ 
  NGC 6215    &    SA(s)c    & A &20.5&1.9&12&-20.1&10.54& 10.53\\ 
  NGC 6300    &    SB(rs)b    & B &14.3&5.2&10.98&-20.58&10.09& 10.32\\ 
  NGC 6384    &    SAB(r)bc    & AB &26.6&6.3&11.14&-21.7&  -  & 10.72\\ 
  NGC 6753    &    (R)SA(r)b    & A &40.9&2.5&11.97&-21.92&10.89&  -\\ 
  NGC 6782    &    (R)SAB(r)a    & AB &50.8&2.2&11.84&  -  &  -  &  - \\ 
  NGC 6902    &    SA(r)b    & A &35.7&6.8&11.64&-21.83&  -  & 10.33\\ 
  NGC 6907    &    SB(s)bc    & B &43&3.3&11.9&-21.96&11.03&  - \\ 
  NGC 7083    &    SA(s)bc    & A &38.7&3.2&11.87&-21.72&10.45& 10.73\\ 
  NGC 7205    &    SA(s)bc    & A &20.5&3.2&11.55&-20.61&10.07& 10.3\\ 
  NGC 7213    &    SA(s)a    & A &22&2.1&11.01&-21.59&  -  & 10.34\\ 
  NGC 7217    &    (R)SA(r)ab    & A &16&3.6&11.02&-20.9&9.9& 10.34\\ 
  NGC 7412    &    SB(s)b    & B &21.1&4.3&11.88&-20.27&  -  & 10.12\\ 
  NGC 7418    &    SAB(rs)cd    & AB &17.8&3.6&11.65&  -  &10.01& 9.96\\ 
  NGC 7479    &    SB(s)c    & B &32.4&3.9&11.6&-21.7&10.79& 10.64\\ 
  NGC 7552    &    (R')SB(s)ab    & B &19.5&3.5&11.25&-20.88&11.03& 10.25\\ 
  NGC 7713    &    SB(r)d    & B &8.2&4.6&11.51&-18.38&  -  & 9.58\\ 
  NGC 7723    &    SB(r)b    & B &23.7&3.9&11.94&-20.66&  -  & 10.31\\ 
  NGC 7727    &    SAB(s)a pec    & AB &23.3&3.3&11.5&-21.25&  -  & 10.34\\ 
  NGC 7741    &    SB(s)cd    & B &12.3&4.1&11.84&-19.14&  -  & 9.7\\ 
 & & & & & & & & \\
\multicolumn {9}{c} {\small Highly inclined galaxies with $i>$~60$^\circ$ (N=33)} \\
\hline\\
IC 4402    &    SA(s)b sp    & A &22.9&5&12&-20.24&10.05&  - \\ 
  IC 5052    &    SBd sp    & B &6.7&5&11.16&-18.6&  -  & 9.28\\ 
  NGC 0625    &    SB(s)m sp    & B &3.9&5&11.91&-16.61&8.57& 8.73\\ 
  NGC 0779    &    SAB(r)b    & AB &17.3&4.4&11.95&-20.03&  -  & 10.2\\ 
  NGC 0908    &    SA(s)c    & A &17.8&6.1&10.83&-21.07&10.27& 10.51\\ 
  NGC 1003    &    SA(s)cd    & A &10.7&6.3&12&-18.7&  -  & 9.64\\ 
  NGC 1421    &    SAB(rs)bc    & AB &25.5&3.5&11.95&-20.61&10.25& 10.64\\ 
  NGC 1808    &    (R)SAB(s)a    & AB &10.8&7.6&10.74&-20.24&10.71& 10\\ 
  NGC 1964    &    SAB(s)b    & AB &20&6.1&11.58&-20.7&10.09& 10.37\\ 
  NGC 2090    &    SA(rs)c    & A &10.2&6.8&11.99&-18.84&  -  & 9.61\\ 
  NGC 2280    &    SA(s)cd    & A &23.2&6.8&10.9&-21.53&10.13& 10.6\\ 
  NGC 3511    &    SA(s)c    & A &15.5&6.8&11.53&-19.99&9.82& 10.25\\ 
  NGC 3675    &    SA(s)b    & A &12.8&5.8&11&  -  &9.92& 10.13\\ 
  NGC 3705    &    SAB(r)ab    & AB &17&4.6&11.86&-20.08&  -  & 10.25\\ 
  NGC 3877    &    SA(s)c    & A &17&5.1&11.79&-20.16&9.89& 10.29\\ 
  NGC 4062    &    SA(s)c    & A &9.7&4.5&11.9&-18.79&  -  & 9.5\\ 
  NGC 4100    &    (P)SA(rs)bc    & A &17&5.1&11.89&-19.99&10.04& 10.25\\ 
  NGC 4293    &    (R)SB(s)0    & B &17&6.3&11.26&-20.79&  -  & 10.21\\ 
  NGC 4388    &    SA(s)b    & A &16.8&5.6&11.76&-20.11&10& 10.16\\ 
  NGC 4448    &    SB(r)ab    & B &9.7&3.8&12&-18.86&  -  & 9.56\\ 
  NGC 4527    &    SAB(s)bc    & AB &13.5&6.3&11.38&-20.13&10.42& 10.08\\ 
  NGC 4654    &    SAB(rs)cd    & AB &16.8&4.8&11.1&-20.63&10.1& 10.32\\ 
  NGC 4666    &    SABc    & AB &14.1&4.2&11.49&-20.01&10.36& 10.1\\ 
  NGC 4772    &    SA(s)a    & A &16.3&2.8&11.96&-20.02&  -  & 9.7\\ 
  NGC 4818    &    SAB(rs)ab pec    & AB &21.5&3.4&12&-20.55&9.75& 10.46\\ 
  NGC 4856    &    SB(s)0    & B &21.1&3.8&11.49&-21.12&  -  & 10.3\\ 
  NGC 5078    &    SA(s)a sp    & A &27.1&4&12&-21.2&10.5&  - \\ 
  NGC 5161    &    SA(s)c    & A &33.5&6.1&12&-21.42&  -  & 10.64\\ 
  NGC 5448    &    (R)SAB(r)a    & AB &32.6&4&11.93&  -  &  -  & 10.47\\ 
  NGC 7184    &    SB(r)c    & B &34.1&6.1&11.65&-21.81&  -  & 10.73\\ 
  NGC 7582    &    (R')SB(s)ab    & B &17.6&4.5&11.37&-20.61&10.87& 10.26\\ 
  NGC 7606    &    SA(s)b    & A &28.9&5.2&11.51&-21.55&  -  & 10.7\\ 
  NGC 7814    &    SA(s)ab: sp    & AB &16&5.5&11.56&-20.45&  -  & 10.18\\
\enddata
\tablecomments{ Columns are : 
(1) Galaxy name; 
(2) Hubble type from  RC3 (de Vaucouleurs \etal 1991); 
(3) RC3 bar type, which is based on visual inspection of optical images 
and runs as `B'=`strongly barred', `AB'=`weakly barred', and `A'=`unbarred'; 
(4) Distance in Mpc. Most values are  from the NBG  (Tully 1988), 
which assumes a Hubble constant of 75 km~s$^{-1}$~Mpc$^{-1}$. 
Exceptions are  NGC~6753, NGC~ 6782, NGC~5078, NGC~6907, NGC~7814, 
and ESO 142-19, for which distances from RC3 are used; 
(5) $D_{\tiny \rm 25}$ in arcminutes, the diameter of the isophote where the 
$B$ band surface brightness is 25 magnitude arcsecond$^{-2}$. Values are from the 
NBG, except for NGC~6753, NGC~6782, NGC~5078, NGC~6907, NGC~7814, 
and ESO 142-19 where RC3 data are used; 
(6) $B_{\tiny \rm T}$, the total blue magnitude from RC3; 
(7) $M_{\tiny \rm V}$, the absolute $V$ magnitude from RC3;
(8) $L_{\tiny \rm IR}$, the global IR luminosity (8 -- 1000 $\mu$m) in units of
log(\mbox{L$_{\odot}$}), from the IRAS Revised Bright Galaxy Sample
(Sanders et al. 2003); 
(9) $L_{\tiny \rm B}$, the global blue luminosity in units of
log(\mbox{L$_{\odot}$}), from the RC3.
}
\end{deluxetable}

\setcounter{table}{1}
\begin{deluxetable}{lcc}
\tabletypesize{\scriptsize}
\tablewidth{0pt}
\tablecaption{Bar statistics from sample S4 (136 galaxies)}
\tablehead{
\colhead {Band} & 
\colhead {Unbarred} &
\colhead {Barred}\\
}
\startdata
$B$ (observed) & 75~=~55\%& 61~=~45\% \\
$H$ (observed) & 57~=~42\%& 79~=~58\%\\
$B$ (deprojected)& 76~=~56\%& 60~=~44\%\\
$H$ (deprojected) & 54~=~40\%& 82~=~60\%\\
\enddata
\tablecomments{ Columns are : 
(1) Band (observed or deprojected);
(2) Number and fraction of galaxies classified as unbarred; 
(3) Number and fraction of galaxies classified as barred. 
}
\end{deluxetable}

\setcounter{table}{2}
\begin{deluxetable}{lcccccccc}
\tabletypesize{\scriptsize}
\tablewidth{0pt}
\tablecaption{Structural properties of sample S4 (136 galaxies) in B and H}
\tablehead{
\colhead {Galaxy Name} & 
\colhead {$i$} &
\colhead {$PA_{\rm disk}$} &
\colhead { class (B) } &
\colhead { $e_{\rm bar}$ (B)  } &
\colhead { $a_{\rm bar}$ (B) } &
\colhead { class (H)  } &
\colhead { $e_{\rm bar}$ (H)  }&
\colhead { $a_{\rm bar}$ (H) } \\
\colhead{}   & 
\colhead{($^\circ$)  }   & 
\colhead{  }   & 
\colhead{   }   & 
\colhead{   }   & 
\colhead{ (kpc) }   &
\colhead{  }   &
\colhead{  } &
\colhead{ (kpc) }  \\
\colhead {(1)} & \colhead {(2)} & \colhead {(3)} & \colhead {(4)} &
\colhead {(5)} & \colhead {(6)} & \colhead {(7)} & 
\colhead {(8)} & \colhead {(9)} \\
}
\startdata
 IC 0239   &37&171	   &   u   & - & - &      & - & -\\
 IC 4444   &36&77	   &   u   & - & - &   u   & - & - \\
 IC 5325   &38&34	   &   u   & - & - &   b   &0.5&1.5\\
 NGC 0157   &41&43 &   u   & - & - &   u   & - & - \\
 NGC 0210   &49&160   &   b   &0.6&12.3&   b   &0.4&9.7\\
 NGC 0278   &25&161   &   u   & - & - &   u   & - & - \\
 NGC 0289   &35&162  &   u   & - & - &   b   &0.6&2.1\\
 NGC 0428   &47&109   &   u   & - & - &   u   & - & - \\
 NGC 0488   &38&6	    &   u   & - & - &   u   & - & - \\
 NGC 0685   &35&95	 &   b   &0.6&2.0&   b   &0.6&1.2\\
 NGC 0864   &43&31  &   b   &0.7&4.2&   b   &0.6&3.5\\
 NGC 1042   &40&173   &   b   &0.6&4.6&   b   &0.6&4.4\\
 NGC 1058   &13&79  &   u   & - & - &   u   & - & - \\
 NGC 1073   &24&174   &   b   &0.7&4.2&   b   &0.7&4.5\\
 NGC 1084   &39&58   &   u   & - & - &   b   &0.4&6.1\\
 NGC 1087   &52&4	   &   u   & - & - &   u   & - & - \\
 NGC 1187   &30&130  &   b   &0.7&3.5&   b   &0.5&2.9\\
 NGC 1241   &55&151  &   b   &0.6&3.8&   b   &0.6&4.1\\
 NGC 1300   &55&102   &   b   &0.6&8.4&   b   &0.5&8.5\\
 NGC 1302   &22&13   &   b   &0.3&2.9&   b   &0.3&2.9\\
 NGC 1309   &21&64   &   u   & - & - &   u   & - & - \\
 NGC 1317   &29&171   &   b   &0.4&0.6&   b   &0.4&0.6\\
 NGC 1350   &58&2	   &   u   & - & - &   b   &0.5&6.8\\
 NGC 1371   &24&81	   &   u   & - & - &   b   &0.4&1.9\\
 NGC 1385   &47&24&   b   &0.8&2.0&   b   &0.6&1.7\\
 NGC 1493   &21&90	  &   u   & - & - &   b   &0.5&1.2\\
 NGC 1559   &56&61   &   b   &0.8&1.7&   b   &0.5&0.9\\
 NGC 1617   &58&109   &   u   & - & - &   u   & - & - \\
 NGC 1637   &35&31	   &   b   &0.5&1.2&   b   &0.4&1.0\\
 NGC 1703   &30&134  &   u   & - & - &   b   &0.3&1.3\\
 NGC 1792   &50&139   &   u   & - & - &   b   &0.5&4.2\\
 NGC 1832   &48&11   &   b   &0.6&2.5&   b   &0.4&2.2\\
 NGC 2139   &36&154   &   u   & - & - &   u   & - & - \\
 NGC 2196   &45&57   &   u   & - & - &   u   & - & - \\
 NGC 2566   &24&59  &   b   &0.6&4.9&   b   &0.5&4.7\\
 NGC 2775   &24&24  &   u   & - & - &   u   & - & - \\
 NGC 2964   &49&95  &   b   &0.6&2.6&   b   &0.5&2.6\\
 NGC 3166   &56&77   &   u   & - & - &   b   &0.5&3.8\\
 NGC 3169   &55&58&   u   & - & - &   u   & - & - \\
 NGC 3223   &47&117   &   u   & - & - &   u   & - & - \\
 NGC 3227   &55&151   &   u   & - & - &   u   & - & - \\
 NGC 3261   &28&59   &   b   &0.5&5.4&   b   &0.4&3.7\\
 NGC 3275   &21&150   &   b   &0.6&7.6&   b   &0.5&6.6\\
 NGC 3423   &39&35   &   u   & - & - &   u   & - & - \\
 NGC 3504   &8&79	   &   b   &0.6&3.8&   b   &0.6&4.1\\
 NGC 3507   &21&67   &   b   &0.5&2.9&   b   &0.5&2.6\\
 NGC 3513   &43&63	   &   b   &0.8&2.7&   b   &0.7&2.1\\
 NGC 3583   &39&134   &   u   & - & - &   b   &0.5&4.4\\
 NGC 3593   &57&86   &   u   & - & - &   u   & - & - \\
 NGC 3596   &32&81   &   u   & - & - &   u   & - & - \\
 NGC 3646   &56&56  &   u   & - & - &   u   & - & - \\
 NGC 3681   &24&132   &   b   &0.3&1.1&   b   &0.3&0.9\\
 NGC 3684   &47&127   &   u   & - & - &   u   & - & - \\
 NGC 3686   &33&18   &   b   &0.7&2.7&   b   &0.5&2.6\\
 NGC 3726   &52&13   &   b   &0.7&4.0&   b   &0.6&3.8\\
 NGC 3810   &45&17   &   u   & - & - &   u   & - & - \\
 NGC 3885   &58&114  &   u   & - & - &   u   & - & - \\
 NGC 3887   &44&13   &   b   &0.6&3.7&   b   &0.5&3.3\\
 NGC 3893   &48&10	 &   u   & - & - &   b   &0.5&6.0\\
 NGC 3938   &30&13  &   u   & - & - &   u   & - & - \\
 NGC 3949   &16&143 &   u   & - & - &   u   & - & - \\
 NGC 4027   &38&176	  &   u   & - & - &   b   &0.6&1.2\\
 NGC 4030   &43&21	   &   u   & - & - &   u   & - & - \\
 NGC 4051   &30&116   &   b   &0.6&4.2&   b   &0.6&4.8\\
 NGC 4123   &43&121   &   b   &0.6&10.5&   b   &0.6&4.2\\
 NGC 4136   &20&51 &   b   &0.6&1.7&   b   &0.4&0.7\\
 NGC 4145   &57&98   &   b   &0.6&1.4&   b   &0.5&1.7\\
 NGC 4151   &36&2	  &   u   & - & - &   b   &0.5&7.9\\
 NGC 4212   &43&72 &   u   & - & - &   b   &0.4&2.8\\
 NGC 4242   &45&22&   u   & - & - &   b   &0.3&3.0\\
 NGC 4254   &24&59   &   u   & - & - &   u   & - & - \\
 NGC 4303   &30&144   &   b   &0.7&3.3&   b   &0.5&4.5\\
 NGC 4314   &18&38 &   b   &0.6&3.2&   b   &0.6&3.7\\
 NGC 4394   &24&109 &   b   &0.5&4.0&   b   &0.5&3.6\\
 NGC 4414   &44&166   &   u   & - & - &   u   & - & - \\
 NGC 4450   &44&174&   b   &0.4&3.8&   b   &0.4&3.6\\
 NGC 4457   &25&86  &   u   & - & - &   u   & - & - \\
 NGC 4487   &48&72 &   u   & - & - &   b   &0.3&1.0\\
 NGC 4496   &24&65	 &   b   &0.7&2.1&   b   &0.6&0.8\\
 NGC 4504   &55&147 &   u   & - & - &   u   & - & - \\
 NGC 4548   &40&154  &   b   &0.6&6.2&   b   &0.6&5.9\\
 NGC 4571   &33&36  &   u   & - & - &   u   & - & - \\
 NGC 4579   &35&95	  &   b   &0.4&3.9&   b   &0.4&3.7\\
 NGC 4580   &43&161   &   b   &0.6&4.3&   b   &0.3&1.9\\
 NGC 4593   &43&105 &   b   &0.6&13.6&   b   &0.6&12.9\\
 NGC 4618   &25&178  &   u   & - & - &   b   &0.6&0.5\\
 NGC 4643   &34&56	  &   b   &0.5&7.1&   b   &0.5&5.4\\
 NGC 4647   &50&121  &   b   &0.6&2.5&   u   & - & - \\
 NGC 4651   &49&71   &   u   & - & - &   u   & - & - \\
 NGC 4665   &33&17  &   b   &0.3&3.8&   b   &0.4&4.1\\
 NGC 4689   &44&173   &   u   & - & - &   u   & - & - \\
 NGC 4691   &34&27	   &   u   & - & - &   b   &0.7&2.01\\
 NGC 4698   &59&174 &   u   & - & - &   b   &0.5&2.6\\
 NGC 4699   &33&34   &   b   &0.3&1.7&   b   &0.3&1.5\\
 NGC 4775   &18&47 &   u   & - & - &   u   & - & - \\
 NGC 4900   &22&113 &   b   &0.8&5.4&   b   &0.6&1.8\\
 NGC 4902   &16&102   &   b   &0.6&6.8&   b   &0.5&4.7\\
 NGC 4930   &40&52  &   b   &0.5&8.1&   b   &0.4&8.0\\
 NGC 4939   &53&6	&   u   & - & - &   u   & - & - \\
 NGC 4941   &58&13 &   u   & - & - &   u   & - & - \\
 NGC 4995   &47&93   &   b   &0.6&6.3&   b   &0.5&3.7\\
 NGC 5005   &59&59   &   u   & - & - &   u   & - & - \\
 NGC 5054   &52&159  &   u   & - & - &   u   & - & - \\
 NGC 5085   &32&56   &   u   & - & - &   u   & - & - \\
 NGC 5101   &23&65  &   b   &0.5&7.3&   b   &0.5&6.8\\
 NGC 5121   &48&57 &   u   & - & - &   u   & - & - \\
 NGC 5247   &36&36  &   u   & - & - &   u   & - & - \\
 NGC 5334   &41&11 &   b   &0.6&3.0&   b   &0.5&1.8\\
 NGC 5371   &40&31  &   b   &0.5&6.5&   b   &0.4&19.7\\
 NGC 5427   &38&11   &   u   & - & - &   b   &0.5&4.8\\
 NGC 5483   &34&50   &   u   & - & - &   u   & - & - \\
 NGC 5676   &59&50  &   u   & - & - &   u   & - & - \\
 NGC 5701   &24&43	  &   b   &0.4&5.8&   b   &0.4&5.3\\
 NGC 5713   &32&1	 &   u   & - & - &   b   &0.6&3.5\\
 NGC 5850   &29&178 &   b   &0.7&12.1&   b   &0.6&10.6\\
 NGC 5921   &46&130   &   b   &0.7&8.5&   b   &0.6&7.6\\
 NGC 5962   &42&109 &   u   & - & - &   u   & - & - \\
 NGC 6215   &44&43&   u   & - & - &   b   &0.5&1.8\\
 NGC 6300   &38&109   &   b   &0.7&3.0&   b   &0.5&2.8\\
 NGC 6384   &55&27   &   u   & - & - &   u   & - & - \\
 NGC 6753   &30&25  &   u   & - & - &   u   & - & - \\
 NGC 6782   &28&36   &   b   &0.5&6.7&   b   &0.4&6.4\\
 NGC 6902   &22&162   &   u   & - & - &   b   &0.3&3.3\\
 NGC 6907   &51&65  &   u   & - & - &   u   & - & - \\
 NGC 7083   &54&8	&   u   & - & - &   u   & - & - \\
 NGC 7205   &58&65  &   u   & - & - &   u   & - & - \\
 NGC 7213   &18&179 &   u   & - & - &   u   & - & - \\
 NGC 7217   &30&136   &   u   & - & - &   u   & - & - \\
 NGC 7412   &52&74   &   b   &0.6&1.8&   b   &0.6&6.6\\
 NGC 7418   &27&91   &   b   &0.7&2.6&   b   &0.6&2.8\\
 NGC 7479   &41&33  &   b   &0.7&8.0&   b   &0.6&8.6\\
 NGC 7552   &23&33  &   b   &0.7&2.0&   b   &0.6&5.3\\
 NGC 7713   &59&166   &   u   & - & - &   u   & - & - \\
 NGC 7723   &34&38   &   b   &0.6&3.2&   b   &0.5&2.3\\
 NGC 7727   &16&64 &   u   & - & - &   u   & - & - \\
 NGC 7741   &40&167 &   b   &0.7&3.2&   b   &0.6&3.0\\

\enddata
\tablecomments{ Columns are : 
(1) Galaxy name; 
(2) Outer disk inclination $i$, calculated from $B$ band ellipse fits before deprojection; 
(3) Outer disk PA, calculated from $B$ band ellipse fits before deprojection; 
(4) $B$ band classification as unbarred (u) or barred (b) from ellipse fits after deprojection;
(5) Bar strength, as characterized by $e_{\rm bar}$, of large-scale bar in $B$ band after deprojection; 
(6) Bar semi-major axis $a_{\rm bar}$ in kpc of large-scale bar in $B$ band after deprojection; 
(7) $H$ band classification as unbarred (u) or barred (b) from ellipse fits after deprojection;
(8) Bar strength, as characterized by $e_{\rm bar}$, of large-scale bar in $H$ band after deprojection; 
(9) Bar semi-major axis $a_{\rm bar}$ in kpc of large-scale bar in $H$ band after deprojection. 
}
\end{deluxetable}


\begin{thebibliography}{}

\bibitem[]{} 
Abraham, R.~G., Merrifield, M.~R., Ellis, R.~S., Tanvir, N.~R., \& Brinchmann, J.\ 1999, 
\mnras, 308, 569 

\bibitem[]{} 
Abraham, R.~G., \& Merrifield, M.~R.\ 2000, \aj, 120, 2835 
 
\bibitem[[]{} 
Aguerri, J.~A.~L., Debattista, V.~P., \& Corsini, E.~M.\ 2003, \mnras, 338, 465
                                                                                                        


\bibitem[]{} 
Athanassoula, E.\ 1992a,\mnras, 259, 328  

\bibitem[]{} 
Athanassoula, E.\ 1992b, \mnras, 259, 345 



\bibitem[]{}
 Athanassoula, E; 2002, ApJL, 569, 83


\bibitem[]{} 
Athanassoula, E.\ 2003,\mnras, 341, 1179 

\bibitem []{} 
Athanassoula, E.\ 2005, \mnras, 358, 1477

\bibitem[]{} 
Athanassoula, E.,Lambert, J.~C., \& Dehnen, W.\ 2005, \mnras, 363, 496


\bibitem[]{} 
Barazza, F.~D., Jogee, S., \& Marinova, I.\ 2006, IAU Symposium, 235 (astroph/0610561)


\bibitem[]{} 
Berentzen, I., Shlosman, I., \& Jogee, S.\ 2006, \apj, 637, 582

\bibitem[]{} 
Berentzen, I., \& Shlosman, I.\ 2006, \apj, 648, 807 


\bibitem[]{} 
Block, D.~L., Bournaud, F., Combes, F., Puerari, I., \& Buta, R.\ 2002, \aap, 394, L35 

\bibitem[]{}
Bournaud, F., \& Combes, F.\ 2002, \aap, 392, 83 

\bibitem[]{} 
Bournaud, F., Combes, F., \& Semelin, B.\ 2005, \mnras, 364, L18 


\bibitem[]{} 
Bureau, M., \& 
Athanassoula, E.\ 2005, \apj, 626, 159 


\bibitem[]{} 
Buta, R., Block, D.~L., \& Knapen, J.~H.\ 2003, \aj, 126, 1148 

\bibitem[]{} 
Buta, R., Laurikainen, E., \& Salo, H.\ 2004, \aj, 127, 279

\bibitem[]{} Buta, R., Vasylyev, S., 
Salo, H., \& Laurikainen, E.\ 2005, \aj, 130, 506 






\bibitem[]{} 
Combes, F., Debbasch, F., Friedli, D., \& Pfenniger, D.\ 1990, \aap, 233, 82
 

\bibitem[]{} 
Curir, A., Mazzei, P., \& Murante, G.\ 2006, \aap, 447, 453 


\bibitem[]{} 
Debattista, V.~P., \& Sellwood, J.~A.\ 1998, \apjl, 493, L5 

\bibitem[]{}
Debattista, V.~P., \& Sellwood, J.~A.\ 2000, \apj, 543, 704

\bibitem[]{} 
Debattista, V.~P., Corsini, E.~M., \& Aguerri, J.~A.~L.\ 2002, \mnras, 332, 65
                                                                                                        


\bibitem[]{} 
Debattista, V.~P., Mayer, L., Carollo, C.~M., Moore, B., Wadsley, J., \& Quinn, T.\ 2006, 
\apj, 645, 209 
                                 
\bibitem{}
de Vaucouleurs, G., de Vaucouleurs, A., Corwin Jr., H. G., 
Buta, R. J., Paturel, G., \& Fouque, P. 1991, Third Reference Catalogue 
of Bright Galaxies  (New York: Springer) (RC3)

\bibitem[]{} 
Dubinski, J.\ 1994, \apj, 431, 617


\bibitem[]{} 
Elmegreen, B.~G., \& Elmegreen, D.~M.\ 1985, \apj, 288, 438 

\bibitem[]{}
Elmegreen, B.~G.\ 1994, \apjl, 425, L73 

\bibitem[]{} Elmegreen, B.~G., 
Elmegreen, D.~M., Chromey, F.~R., Hasselbacher, D.~A., \& Bissell, B.~A.\ 
1996, \aj, 111, 2233 

\bibitem{}
Elmegreen, B.~G., Elmegreen, D.~M., \& Hirst, A.~C.\ 2004, \apj, 612,
191 


\bibitem[]{} 
Erwin, P.\ 2005, \mnras, 364, 
283 

\bibitem[]{} 
Eskridge, P.~B., et al.\ 2000, \aj, 119, 536

\bibitem[]{} 
Eskridge, P.~B., et al.\ 2002, \apjs, 143, 73

\bibitem[]{} 
Friedli, D., Wozniak, H., Rieke, M., Martinet, L., \& Bratschi, P.\ 1996, \aaps, 118, 461 

\bibitem[]{} 
Giavalisco, M., et al.\ 2004, \apjl, 600, L93



\bibitem[]{} 
Hunt, L.~K., \& Malkan, M.~A.\ 1999, \apj, 516, 660 



\bibitem[]{} 
Jedrzejewski, R.~I.\  1987, \mnras, 226, 747


\bibitem{}
Jogee, S.  1999, Ph.D. thesis, Yale University 

\bibitem[]{} 
Jogee, S., Kenney, J.~D.~P., \& Smith, B.~J.\ 1999, \apj, 526, 665 


\bibitem{}
Jogee, S., Knapen, J. H.,  Laine, S., Shlosman, I., Scoville,
N. Z., \&  Englmaier, P. 2002a, ApJL, 570, L55 

\bibitem{}
Jogee, S., Shlosman, I., Laine, S.,   Knapen, J. H., Englmaier, 
P., Scoville, N. Z., \& Wilson, C. D. 2002b, ApJ


\bibitem{}
Jogee, S.,  Barazza, F.,  Rix, H.-W., Shlosman, I. \etal  2004a, ApJl, 615, 
L105 



\bibitem{}
Jogee, S., Scoville, N., \& Kenney, J.~D.~P.\ 2005, \apj, 630, 837 

\bibitem{}
Jogee, S.,  2006, Lecture Notes in Physics, Vol. 693, "AGN Physics 
on All Scales", Eds. D. Alloin, R. Johnson, \& P. Lira (Springer:
Berlin Heidelberg New  York), Vol 93, Chapter 6, p 143 (astro-ph/0408383).

\bibitem[]{} 
Kazantzidis, S., Kravtsov, A.~V., Zentner, A.~R., Allgood, B., Nagai, D., \& Moore, B.\ 
2004, \apjl, 611, L73 


\bibitem[]{}
Knapen, J. H., Beckman, J. E., Heller, C. H., Shlosman, I., \& De Jong, R. S. 1995, ApJ, 454, 623



\bibitem[]{} 
Knapen, J.~H., Shlosman, I., \& Peletier, R.~F.\ 2000, \apj, 529, 93






\bibitem{}
Kormendy, J. 1993, in  Proceedings of IAU Symposium 153, Galactic bulges, 
ed. H. DeJonghe and H. J. Habing,  (Dordrecht: Kluwer), 209

\bibitem{}
Kormendy, J., \& Kennicutt, R.~C.\ 2004, \araa, 42, 603





\bibitem[]{} 
Laine, S., Shlosman, I., Knapen, J.~H., \& Peletier, R.~F.\ 2002, \apj, 567, 97 

\bibitem[]{} 
Laurikainen, E., Salo, H., \& Rautiainen, P.\ 2002, \mnras, 331, 880 

\bibitem[]{} 
Laurikainen, E., Salo, H., \& Buta, R.\ 2004, \apj, 607, 103




\bibitem[]{} 
Lisker, T., Debattista, V.~P., Ferreras, I., \& Erwin, P.\ 2006, \mnras, 370, 477
                                                                                                        


\bibitem[]{} 
Martin, P.\ 1995, \aj, 109, 2428

\bibitem[]{} 
Martinez-Valpuesta, I., Shlosman, I., \& Heller, C.\ 2006, \apj, 637, 214 

\bibitem{} 
Menendez-Delmestre, K. et al., 2004, in ``Penetrating Bars Through Masks of Cosmic Dust: The Hubble Tuning Fork Strikes a New Note'', Eds. D. Block, I. Puerari, K. Freeman, R. Groess, \& E. Block (Springer), pp.787-8

\bibitem{} 
Menendez-Delmestre, K. et al. 2006, ApJ, accepted


\bibitem[]{} 
Merrifield, M.~R., \& Kuijken, K.\ 1995, \mnras, 274, 933
                                                                                                                                                                                                               

\bibitem[]{} 
Mulchaey, J.~S., \& Regan, M.~W.\ 1997, \apjl, 482, L135 

 









\bibitem[]{} 
Pfenniger, D., \& Norman, C.\ 1990, \apj, 363, 391 

\bibitem[O'Neill \& Dubinski(2003)]{2003MNRAS.346..251O} 
O'Neill, J.~K., \& Dubinski, J.\ 2003, \mnras, 346, 251





\bibitem[]{} 
Rix, H.-W., et al.\ 2004, \apjs, 152, 163 

\bibitem[]{}
Regan, M. W., Vogel, S. N., \& Teuben, P. J.  1997, ApJ, 482, L143




\bibitem[]{}
Scoville \etal  2006, ApJS, submitted (astro-ph/0612306)


\bibitem[]{} Shen, J., \& 
Sellwood, J.~A.\ 2004, \apj, 604, 614 

\bibitem[]{} 
Sheth, K., Regan, M.~W., Scoville, N.~Z., \& Strubbe, L.~E.\ 2003, \apjl, 592, L13 



\bibitem[]{} 
Shlosman, I., \& Noguchi, M.\ 1993, \apj, 414, 474


\bibitem[]{} 
Tran, H., et al. 2003, ApJ, 585, 750
                                                                                                        
\bibitem[]{} 
Tully, R.~B.\ 1988, Cambridge and New York, Cambridge University Press, 1988, 221 p., (NBG)




\bibitem[]{} 
Weinberg, M.~D.\ 1985, \mnras, 213, 451


\bibitem[]{} 
Whyte, L.~F., Abraham, R.~G., Merrifield, M.~R., 
Eskridge, P.~B., Frogel, J.~A., \& Pogge, R.~W.\ 2002, \mnras, 336, 1281 

\bibitem[]{} 
Wolf, C., et al.\ 2004, \aap, 421, 913 

\bibitem[]{} 
Wozniak, H., Friedli, D., Martinet, L., Martin, P., \& Bratschi, P.\ 1995, \aaps, 111, 115

\bibitem[]{} 
Zheng, X.~Z., Hammer, F., Flores, H., Ass{\'e}mat, F., \& Rawat, A.\ 2005, \aap, 435, 507 



\end{thebibliography}
\end{document}